\documentclass[aps,prb,twocolumn,amsfonts,amssymb,amsmath,floatfix,floats,twoside,fleqn]{revtex4}
\usepackage{graphicx}
\usepackage{bm}

\begin{document}
\title{The \emph{1D} Hubbard model within the Composite Operator Method}
\author{Adolfo Avella}
\email[E-mail: ]{avella@sa.infn.it}
\author{Ferdinando Mancini}
\affiliation{Dipartimento di Fisica ``E.R.~Caianiello'' -- Unit\`a
di Ricerca INFM di Salerno
\\ Universit\`a degli Studi di Salerno, 84081 Baronissi (SA),
Italy}
\author{Maria del Mar S\'anchez-L\'opez}
\affiliation{Departamento de Ciencia y Tecnologi\'a de Materiales
\\ Universidad Miguel Hern\'andez, 03202 Elche (Alicante), Spain}

\date{\today}

\begin{abstract}
Although effective for two dimensional (\emph{2D}) systems, some
approximations may fail in describing the properties of
one-dimensional (\emph{1D}) models, which belong to a different
universality class. In this paper, we analyze the adequacy of the
Composite Operator Method (\emph{COM}), which provides a good
description of many features of \emph{2D} strongly correlated
systems, in grasping the physics of \emph{1D} models. To this
purpose, the \emph{1D} Hubbard model is studied within the
framework of the \emph{COM} by considering a two-pole
approximation and a paramagnetic ground state. The local,
thermodynamic and single-particle properties, the correlation
functions and susceptibilities are calculated in the case of half
filling and arbitrary filling. The results are compared with those
obtained by the Bethe Ansatz (\emph{BA}) as well as by other
numerical and analytical techniques. The advantages and
limitations of the method are analyzed in detail.
\end{abstract}


\maketitle

\section{Introduction}

\label{Intro}

The physics of interacting electrons confined to \emph{1D} systems
is one of the most interesting fields of research in Condensed
Matter Physics. The reasons are various. On the one hand, the
physics of \emph{1D} systems challenges the standard picture of
interacting electrons in metals, which has the Fermi liquid
(\emph{FL}) theory as its basic cornerstone. As Tomonaga and
Luttinger\cite{Tomonaga:50,Luttinger:63} showed, a strictly
\emph{1D} interacting electron system cannot be described by
\emph{FL} theory. In such a system charge and spin degrees of
freedom merge into collective low-energy excitations that
propagate with different velocities and the quasi-particle
picture, essential to \emph{FL} theory, breaks down. This new
electronic state is called Luttinger liquid (\emph{LL}). Signals
of \emph{LL} behavior can be sought in any physical realization of
\emph{1D} electronic systems. Synthetic organic metals like the
Bechgaard salts are probably the best candidates\cite{Jerome:94}.
These metals have crystal structures consisting of alternating
layers of organic donor molecules like \emph{TMTTF} and
\emph{TMTSF}, and inorganic anions such as $PF_{6}$, $SbF_{6}$ or
$Br$. Stacking planar molecules yield an overlap of the molecular
orbitals that is greatest along the stacks and weaker between
them, thus producing quasi-\emph{1D} conductors. Recent optical
measurements in the metallic state of various Bechgaard salts have
shown consistency with \emph{LL} behavior\cite{Vescoli:98}. On the
other hand, the Bechgaard salts are subject of intensive
studies\cite{Jerome:82,Jerome:91} because they have a rich phase
diagram, with antiferromagnetic, spin-Peierls, spin-density wave
and superconducting ground states. In particular, the
superconducting state shows some similarities to that of
high-$T_{c}$ cuprates (high anisotropic conductivity, large and
anisotropic critical field\cite{Welp:89} and short coherence
length\cite{Kwok:90}). Also, the interplay between
antiferromagnetic and superconducting ground states and the strong
sensitivity of $T_{c}$ to non-magnetic impurities indicate an
unconventional superconducting mechanism that still remains to be
determined\cite{Jerome:82,Jerome:91,Williams:91}. All these
properties are undoubtedly a strong motivation for better
understanding the physics of interacting electrons in \emph{1D}
systems. One of the most suitable Hamiltonian to consider for this
purpose is the \emph{1D} Hubbard model\cite{Hubbard:63}. This
Hamiltonian is exactly integrable by means of the
\emph{BA}\cite{Yang:67}. In this way, many properties are known
exactly within the numerics needed in the case of arbitrary
particle density and/or finite temperature. Namely, many
ground-state properties\cite{Lieb:68,Shiba:72} (total energy,
local magnetization, magnetic susceptibility, etc.), charge and
spin excitation spectra\cite{Ovchinnikov:70}, and some
thermodynamic properties\cite{Kawakami:89,Usuki:90,Carmelo:88} can
be exactly computed. However, the \emph{BA} does not provide a
complete framework for describing the physics of the \emph{1D}
Hubbard model since many properties, like the correlation and
spectral functions, cannot be evaluated from the \emph{BA} wave
function except for some limiting cases\cite{Penc:95,Penc:96}
(infinite interaction, static case, half filling). Therefore, to
compute these quantities, which are among the most relevant ones
for describing real materials and getting a complete overview, we
must consider other approaches. Bosonization techniques, conformal
field theory and quantum transfer matrix (\emph{qtm}) and string
theory investigations are analytic methods often used for
\emph{1D} models. They permit to compute key quantities like
correlation functions, scaling relations between their exponents
and the velocities of spin and charge collective modes, but also
thermodynamic quantities like the specific heat and the charge and
spin susceptibilities \cite{Voit:95,Juttner:98,Deguchi:00}.
However, these methods need lengthy and complex calculations. The
numerical
techniques\cite{Shiba:72a,Hirsch:83,Imada:89,Sorella:90,Xu:92,Preuss:94,Schulte:96,Bedurftig:98},
instead, are limited by the small size of clusters and the
impossibility of reaching very low temperatures. We are thus
interested in analyzing the adequacy of a simpler analytical
calculation scheme for describing the physics of correlated
electrons in \emph{1D} models. This method, called the Composite
Operator Method (\emph{COM}), is based on the choice of an
appropriate combination of standard fermionic field operators as
basis for describing the excitations of the system. The properties
of the composite fields are self-consistently determined through
the equations of motion, and the parameters that arise
(\emph{internal} parameters) are used to fix the representation
where the dynamics is realized\cite{Mancini:00}. This procedure
recovers symmetries that are usually badly violated by other
approaches\cite{Avella:98} and provides a good description of many
features of strongly correlated systems; it is in excellent
agreement with numerical simulations on the local and integrated
quantities\cite{Mancini:95,Mancini:95a,Mancini:95b,Fiorentino:00a}
and explains successfully some anomalous
thermodynamic\cite{Mancini:97,Avella:98c} and magnetic
behaviors\cite{Mancini:98c,Avella:98d} observed in high-$T_{c}$
cuprate superconductors. Nevertheless, approximations adequate in
higher dimensions can fail when applied to \emph{1D} systems. The
\emph{BA} provides a useful test for any approximate method. In
this paper we study the adequacy of \emph{COM} to describe the
physics of the \emph{1D} Hubbard model. We evaluate the local,
thermodynamic and single-particle properties, the correlation
functions and the susceptibilities. Our results are compared to
the \emph{BA} ones, whenever available. We also discuss the
agreement with other analytical and numerical techniques, in
particular, the Renormalization Group (\emph{RG}) and the quantum
Monte Carlo (\emph{qMC}). The advantages and limitations of the
\emph{COM} are discussed. Some preliminary results have already
been published in Refs.~\onlinecite{Avella:98e} and \onlinecite
{Sanchez:99a}; the present work provides an exhaustive overview of
the application of the \emph{COM} to the \emph{1D} Hubbard model.

The paper is structured as follows. In Sec.~\ref{Method}, the
framework of the \emph{COM} for the \emph{1D} Hubbard model is
extensively described. The model is presented, the basis chosen,
the solution given and many physical quantities addressed. In
Sec.~\ref{Results}, the results for half filling and arbitrary
filling are analyzed separately. Special attention is devoted to
the case of quarter filling since, together with the half-filled
case, it is believed to be the scenario for the Bechgaard
salts\cite{Vescoli:98}. Finally, in Sec.~\ref{Conclusions}, some
conclusions are given. In the Appendix, the two-pole approximation
scheme is reported in some detail.

\section{The Method}

\label{Method}

\subsection{The Model}

The \emph{1D} Hubbard model is described by the following
Hamiltonian:
\begin{equation}
H=\sum_{ij}\left[ t_{ij}-\mu\,\delta_{ij}\right]
\,c^{\dagger}\left( i\right) \,c\left(  j\right)
+U\sum_{i}n_{\uparrow}\left(  i\right) \,n_{\downarrow}\left(
i\right)
\end{equation}
where $c^{\dagger}\left(  i\right)  =\left(
c_{\uparrow}^{\dagger}\left( i\right)
,c_{\downarrow}^{\dagger}\left(  i\right)  \right)  $ is the
creation electron operator at the site $i$ in spinor notation,
$n_{\sigma }\left( i\right)
=c_{\sigma}^{\dagger}(i)\,c_{\sigma}(i)$ is the charge density
operator for the spin $\sigma$, $\mu$ is the chemical potential
introduced to control the particle density (i.e., the filling) $n$
and $U$ is the intrasite Coulomb interaction. The hopping matrix
is given by
\begin{equation}
t_{ij}=-2t\frac{1}{N}\sum_{k}e^{ik\left(  i-j\right)  }\cos k
\end{equation}
where unitary lattice constant and only nearest neighbors are
considered. We have fixed the energy scale in such a way that
$t_{ii}=0$. Hereafter, any energy will be presented in units of
$t$ and we will consider $\hbar =k_{{\rm B}}=1$.

\subsection{The Basis}

In the case of the Hubbard model, a natural choice for the
operatorial basis is the Hubbard doublet
$\Psi^{\dagger}(i)=(\xi^{\dagger}(i),\eta^{\dagger}(i)) $, where
\begin{equation}
\xi^{\dagger}(i)=c^{\dagger}(i)\,[1-n(i)]\qquad\eta^{\dagger}(i)=c^{\dagger
}(i)\,n(i)\label{Basis}
\end{equation}
These operators describe the atomic transitions at the site $i$
(i.e., the transitions $n=0\leftrightarrow1$ and
$n=1\leftrightarrow2$, respectively). We have
\begin{align}
&  \xi_{\sigma}^{\dagger}\left|  0\right\rangle =\left|
\sigma\right\rangle
\qquad\xi_{\sigma}^{\dagger}\left\{\begin{array} [c]{l}
\left|  \alpha\right\rangle \\
\left|  \uparrow\downarrow\right\rangle
\end{array}
\right.  =0\nonumber\\
&  \eta_{\sigma}^{\dagger}\left|  \bar{\sigma}\right\rangle
=(-)^{\sigma +1}\left|  \uparrow\downarrow\right\rangle
\qquad\eta_{\sigma}^{\dagger }\left\{\begin{array} [c]{l}
\left|  0\right\rangle \\
\left|  \sigma\right\rangle \\
\left|  \uparrow\downarrow\right\rangle
\end{array}
\right.  =0
\end{align}
where $\alpha,\sigma=\uparrow(1)$ or $\downarrow(2)$ and \{$\left|
0\right\rangle $, $\left|  \sigma\right\rangle $, $\left| \uparrow
\downarrow\right\rangle$\} is the vectorial basis on the single
site.

\subsection{The Green's function and the \emph{COM} solution}

Considering a two-pole approximation\cite{Mancini:98b} (see
App.~\ref{two-pole}) and a paramagnetic ground state, the Fourier
transform of the single-particle retarded thermal Green's function
$G\left(  i,j\right) =\left\langle {\cal
R}\left\{\Psi(i),\Psi^{\dagger}(j)\right\} \right\rangle $ may be
written as
\begin{equation}
G\left(  k,\omega\right) =\sum_{i=1}^{2}\frac{\sigma^{(i)}\left(
k\right) }{\omega-E_{i}\left(  k\right) +i\varepsilon}\label{GF}
\end{equation}
where the spectral functions $\sigma^{(i)}(k)$ are given by
\begin{align}
\sigma_{11}^{(i)}(k) &  =I_{11}\frac{2Q(k)+(-)^{i+1}g(k)}{4Q(k)}\\
\sigma_{12}^{(i)}(k) &  =\sigma_{21}^{(i)}(k)=(-)^{i+1}\frac{m(k)}{2Q(k)}\\
\sigma_{22}^{(i)}(k) &  =I_{22}\frac{2Q(k)+(-)^{i}g(k)}{4Q(k)}
\end{align}
and $E_{i}(k)=R(k)-(-)^{i}\,Q(k)$ are the energy spectra, with
\begin{align}
R(k) &  =\frac{1}{2}U-\mu-2t\cos k-\frac{m(k)}{2\,I_{11}I_{22}}\\
Q(k) &  =\frac{1}{2}\sqrt{g^{2}(k)+4\frac{m^{2}(k)}{I_{11}I_{22}}}\\
g(k) &  =(1-n)\frac{m(k)}{I_{11}I_{22}}-U\\
m(k) &  =2t\left[  \Delta+\left(  p-I_{22}\right)  \cos k\right]
\label{lastGF}
\end{align}
$I_{11}=1-n/2$ and $I_{22}=n/2$ are the diagonal elements of the
normalization matrix (see App.~\ref{two-pole}). As we can see from
Eqs.~(\ref{GF})-(\ref{lastGF}), the Green's function depends on
the model parameters $t $ and $U$, the \emph{external} parameters
$n$ and $T$ (temperature), and three \emph{internal} parameters:
the chemical potential $\mu$, $\Delta$ and $p$. The latter two
parameters have the following expressions
\begin{align}
\Delta &
=\langle\xi^{\alpha}(i)\,\xi^{\dagger}(i)\rangle-\langle\eta
^{\alpha}(i)\,\eta^{\dagger}(i)\rangle\\
p &  =\frac{1}{4}\langle
n_{\mu}^{\alpha}(i)\,n_{\mu}(i)\rangle-\langle
(c_{\uparrow}(i)\,c_{\downarrow}(i))^{\alpha}c_{\downarrow}^{\dagger
}(i)\,c_{\uparrow}^{\dagger}(i)\rangle
\end{align}
and they are related\cite{Avella:98} to the difference between the
hopping amplitudes within the two Hubbard subbands ($\Delta$) and
the intersite charge, spin and pair correlation functions ($p$).
The superscript $\alpha$ indicates the projection on the first
neighbor sites
\begin{equation}
\phi^{\alpha}(i)=\sum_{j}\alpha_{ij}\,\phi(j)
\end{equation}
$n_{\mu}(i)=c^{\dagger}(i)\,\sigma_{\mu}\,c(i)$ are the charge
($\mu=0$) and spin ($\mu=1,2,3$) density operators, where
$\sigma_{\mu}=\left(  1, \vec{\sigma}\right)  $,
$\sigma^{\mu}=\left(  -1,\vec{\sigma}\right)  $ and $\vec{\sigma}$
are the Pauli matrices. The main effect of the internal parameters
$\Delta$ and $p$ on the bands $E_{i}(k)$ is a uniform shift and a
bandwidth renormalization, respectively. Depending on how these
internal parameters are fixed\cite{Avella:98}, very different
results are obtained. In the \emph{COM} they are determined by
solving the following system of coupled self-consistent equations,
\begin{equation}
\left\{\begin{array} [c]{l}
n=2\left(  1-C_{11}-C_{22}\right) \\
\Delta=C_{11}^{\alpha}-C_{22}^{\alpha}\\
C_{12}=0
\end{array}
\right. \label{self}
\end{equation}
with $C_{\gamma\delta}=\left\langle
\Psi_{\gamma}(i)\Psi_{\delta}^{\dagger }(i)\right\rangle $ and
$C_{\gamma\delta}^{\alpha}=\left\langle \Psi_{\gamma
}^{\alpha}(i)\Psi_{\delta}^{\dagger}(i)\right\rangle $. The first
two equations come from the existing relations between the
parameters $n$ and $\Delta$ and the Green's function matrix
elements, whereas the third equation has been chosen in order to
satisfy the Pauli principle at the level of matrix
elements\cite{Mancini:00,Avella:98}. This request, which fixes the
proper representation of the Hilbert space, naturally implies the
fulfillment of the constrains coming from the particle-hole
sysmmetry\cite{Avella:98}; i.e.,
\begin{align}
\mu(2-n) & =U-\mu(n)\\
\Delta(2-n) & =-\Delta(n)\\
p(2-n) & =1-n+p(n)\label{pa-ho}
\end{align}
Let us note that, at half filling, the third self-consistent
equation in (\ref{self}) is identically satisfied and the $p$
parameter must be calculated by analytic continuation. In this
case we have $\mu=U/2$, $\Delta=0$ and the energy spectra
($i=1,2$) have the following expression
\begin{equation}
E_{i}(k)=-4t\,p\cos k-\frac{1}{2}(-)^{i}\sqrt{U^{2}+16t^{2}\left(
2p-1\right) ^{2}\cos^{2}k}
\end{equation}

\subsection{The physical properties}

Within this calculation scheme the evaluation of the physical
properties is straightforward once the internal parameters are
determined. In the following we will describe how the relevant
quantities can be computed.

\subsubsection{The local quantities}

The chemical potential $\mu$ is one of the outputs of
Eqs.~(\ref{self}). In the non-interacting case, $\mu$ is
detetmined as a function of the particle density $n$ and of the
temperature $T$ by means of the equation
\begin{equation}
n=1-\frac{1}{\pi}\int_{0}^{\pi}\left\{ 1-2\,f_{{\rm F}}\left[
E(k)\right] \right\}  dk
\end{equation}
where
\begin{equation}
E(k)=-\mu-2t\cos k\label{bandU0}
\end{equation}
is the non-interacting energy spectrum and $f_{{\rm F}}(\omega)$
is the Fermi function. At zero temperature, the previous equation
can be solved analitically and gives
\begin{equation}
\mu=-2t\cos\left(  \frac{\pi}{2}n\right)  \label{munonint}
\end{equation}

The internal energy per site $E$ can be calculated as the thermal
average of the Hamiltonian and is given by
\begin{equation}
E=4t\sum_{\gamma\delta}C_{\gamma\delta}^{\alpha}+U\,D\label{EH}
\end{equation}
where $D=\left\langle
n_{\uparrow}(i)\,n_{\downarrow}(i)\right\rangle $ is the double
occupancy. In the insulating phase, at zero temperature and half
filling, the previous equation assumes the following expression
\begin{align}
E  & =\frac{U}{4}+\frac{\sqrt{U^{2}+a\,t^{2}}}{\pi\left(
2p-1\right)  }
{\cal E}\left(  \sqrt{\frac{a\,t^{2}}{U^{2}+a\,t^{2}}}\right)  \nonumber\\
& -\frac{U^{2}\left(  2p+1\right)  }{2\pi\left(  2p-1\right) \sqrt
{U^{2}+a\,t^{2}}}{\cal K}\left(
\sqrt{\frac{a\,t^{2}}{U^{2}+a\,t^{2}}}\right) \label{En1T0COM}
\end{align}
where $a=16\left(  2p-1\right)  ^{2}$, ${\cal K}(x)$ and ${\cal
E}(x)$ are the complete elliptic integrals of first and second
kind, respectively. In the non-interacting case, we have
\begin{equation}
E=-4t\frac{1}{\pi}\int_{0}^{\pi}f_{{\rm F}}\left[  E(k)\right]
\cos k\,dk
\end{equation}
We recall that, at half filling, the Bethe Ansatz result of Lieb
and Wu\cite{Lieb:68} reads as
\begin{equation}
E=-4t\int_{0}^{\infty}\frac{J_{0}(x)J_{1}(x)}{x\left(
1+e^{x\,U/2t}\right) }\,dx\label{EBA}
\end{equation}
where $J_{n}(x)$ is the Bessel function of order $n$. The double
occupancy $D$ can be obtained as
\begin{equation}
D=\frac{n}{2}-C_{22}\label{DH}
\end{equation}
In the non-interacting case, we have $D=\frac{n^{2}}{4}$. The
local magnetization $L_{0}=\frac{1}{4}\left\langle
n_{3}(i)\,n_{3}(i)\right\rangle $ can be computed by
\begin{equation}
L_{0}=\frac{3}{2}\left(  \frac{n}{2}-D\right)  \label{L0}
\end{equation}
In the non-interacting case, we have $L_{0}=\frac{3}{4}n\left(
1-\frac{n} {2}\right)  $.

\subsubsection{The thermodynamics}

The thermodynamic properties can be computed through the
appropriate integrals of the chemical potential and its
derivatives. In particular, the specific heat $C$ can be obtained
as
\begin{equation}
C(U,T,n)=-T\int_{0}^{n}\,\frac{\partial^{2}}{\partial T^{2}} \mu
(U,T,n')\,dn'\label{cv}
\end{equation}
where we have made use of the thermodynamic relation $E=F+TS$,
with $F$ and $S$ the Helmholtz free energy and the entropy per
site, respectively, and we have exploited the following
expressions
\begin{align}
F(U,T,n)  &  =\int_{0}^{n}\mu(U,T,n')\,dn'\label{FT}\\
S(U,T,n)  &  =-\int_{0}^{n}\frac{\partial}{\partial
T}\mu(U,T,n')\,dn'\label{ST}
\end{align}
which give
\begin{equation}
E(U,T,n)=\int_{0}^{n}\left[ \mu(U,T,n')-T\frac{\partial}{\partial
T} \mu(U,T,n')\right]  dn'\label{ET}
\end{equation}
As it was shown for the \emph{2D} case\cite{Mancini:97}, the
temperature derivatives of the chemical potential can be expressed
in terms of the internal parameters. It remains clear that, once
the self-consistent equations (\ref{self}) are solved, there exist
two ways to calculate the physical quantities. On the one hand, we
can exploit, whenever is possible, the relations with the Green's
function matrix elements; on the other hand, we can use the
relations between the conjugate variables and the Helmholtz free
energy computed through the chemical potential. The energy per
site is a clear example of these two ways of calculation, since it
can be computed by means of both Eqs.~(\ref{EH}) and (\ref{ET}).
Another example is the double occupancy, which can be calculated
by using Eq.~(\ref{DH}) and as
\begin{equation}
D(U,T,n)=\frac{\partial}{\partial
U}F(U,T,n)=\int_{0}^{n}\frac{\partial}{\partial U}
\mu(U,T,n')\,dn'
\end{equation}
Obviously, the exact solution of the model gives identical results
whichever way we choose. On the contrary, any analytical
approximation can receive different results because the first way
mainly exploits the computation of two-particle static correlation
functions while the second way is based on the value of the
chemical potential, which can be computed by using only
one-particle static correlation functions; in this case, for
particles, we intend the original $c$ electrons. Another procedure
to compute the Helmholtz free energy $F$ and the entropy $S$
exploits the relations $E=F+TS$ and $S=- \frac{\partial
F}{\partial T}$. Using the latter, we can rewrite the former as
follows
\begin{equation}
-\frac{E}{T^{2}}=\frac{\partial}{\partial T}\left(
\frac{F}{T}\right)
\end{equation}
and obtain for $\frac{T^{*}}{T}\ll1$
\begin{align}
F(n,T,U) & =E(n,T^{*},U)\nonumber\\
& +T\int_{T^{*}}^{T}\frac{E(n,T^{*},U)-E(n,T',U)}{{T'}^{2} }\,dT'
\end{align}
where the value of the internal energy $E$ is given by
Eq.~(\ref{EH}). Hereafter, we will put a subindex $H$ near any
quantity calculated from the matrix elements of the Green's
function and a subindex $T$ near any quantity computed through the
value of the chemical potential.

\subsubsection{The single-particle properties}

The momentum distribution function $n(k)$ is defined by means of
the equation
\begin{equation}
n=\frac{1}{2\pi}\int_{-\pi}^{\pi}n(k)\,dk
\end{equation}
Within the \emph{COM}, it can be computed as follows
\begin{equation}
n(k)=2\left\{f_{{\rm F}}[E_{1}(k)]Z_{1}(k)+f_{{\rm F}}[E_{2}
(k)]Z_{2}(k) \right\}
\end{equation}
where
\begin{equation}
Z_{i}(k)=\frac{1}{2}\left[ 1-(-)^{i}\frac{(1-n)g(k)}{2Q(k)}\right]
\end{equation}
are the weights of the two subbands. It is easy to verify that
$Z_{i}(k)\leq 1$. In the non-interacting case, we have
\begin{equation}
n(k)=2f_{{\rm F}}\left[  E(k)\right]
\end{equation}
and, at zero temperature, the Fermi momentum $k_{{\rm F}}$
(defined by $E(k_{{\rm F}})=0$) assumes the value
$\frac{\pi}{2}n$.

The density of states for the $c$ electrons is given by the
following expression
\begin{equation}
N_{cc}(\omega)=\frac{1}{2\pi}\sum_{i=1}^{2}\sum_{\gamma\delta}\int_{-\pi}
^{\pi}\sigma^{(i)}_{\gamma\delta}\left( k\right) \,\delta\left[
\omega -E_{i}\left( k\right) \right] \,dk
\end{equation}
In the non-interacting case, we have
\begin{equation}
N_{cc}(\omega)=\frac{1}{2\pi\,t}\frac{1}{\sqrt{1-\left(
\frac{\omega+\mu}{2t} \right)  ^{2}}}\theta\left(  1-\left|
\frac{\omega+\mu}{2t}\right|  \right)
\end{equation}
which clearly exhibits the well-known 1D van Hove singularities at
the edges of the band (cfr. Eq.~\ref{bandU0}). In the interacting
case, each singularity splits in two as we have two distinct
subbands.

\subsubsection{The correlation functions and the susceptibilities}

The distribution function $B(r)=\left\langle
c^{\dagger}(r)c(0)\right\rangle $ can be computed as follows
\begin{equation}
B(r)=2\delta_{r,0}-2\sum_{ab}C_{ab}(r)\label{BCOM}
\end{equation}
where $C_{ab}(r)=\left\langle
\Psi_{a}(r)\Psi_{b}^{\dagger}(0)\right\rangle $ is the static
correlation function given by
\begin{equation}
C_{ab}(r)=\frac{1}{2\pi}\sum_{m=1}^{2}\int_{-\pi}^{\pi}e^{i\,k\,r}\left\{1-f_{{\rm
F}}\left[  E_{m}(k)\right]  \right\} \sigma_{ab}^{(m)}
(k)\,dk\label{corr}
\end{equation}
At zero temperature and half filling, Eq.~(\ref{BCOM}) assumes the
following simple expression
\begin{equation}
B(r)=\delta_{r,0}-4t\left(  2p-1\right)
\frac{1}{2\pi}\int_{0}^{\pi} \frac{\cos\left(  k\,r\right)  \cos
k}{\Gamma\left(  k\right)  }dk
\end{equation}
where $\Gamma\left(  k\right)  =\sqrt{U^{2}/4+4t^{2}\left(
2p-1\right) ^{2}\cos^{2}k}$. In the non-interacting case, we have
\begin{equation}
B(r)=\delta_{r,0}-\frac{1}{\pi}\int_{0}^{\pi}\left\{ 1-2f_{{\rm
F}}\left[ E(k)\right]  \right\}  \cos kr\,dk
\end{equation}
which, at zero temperature, becomes
\begin{equation}
B(r)=\frac{2}{\pi\,r}\sin\frac{\pi\,n\,r}{2}
\end{equation}
showing damped oscillations ($r^{-1}$) of wavelength
$\lambda=\frac{4} {n}=\frac{2\pi}{k_{{\rm F}}}$.

By considering the one-loop approximation\cite{Mancini:95b}, the
two-particle Green's functions can be calculated in terms of the
single-particle ones.

The density-density correlation function $N(r)=\left\langle
n(r)\,n(0)\right\rangle $ will be computed as follows
\begin{equation}
N(r)=n^{2}+\frac{n(2-n)}{n-2D}\sum_{a,b,c=1}^{2}I_{aa}^{-1}Q_{abac}
(r)\label{NCOM}
\end{equation}
where
\begin{equation}
Q_{abac}(r)=\left[  I_{ab}(r)-C_{ab}(r)\right]  C_{ac}(r)
\end{equation}
and $I_{ab}$ are the elements of the normalization matrix. It is
worth noting that Eq.~(\ref{NCOM}) satisfies the sum rule
$N(0)=n+2D$; this is a clear manifestation of the conserving
nature, at level of local sum rules, of the one-loop
approximation. In the non-interacting case, the density-density
correlation function reads as
\begin{equation}
N(r)=n^{2}+\delta_{r,0}n-\frac{1}{2}B^{2}(r)
\end{equation}
and shows damped oscillations ($r^{-2}$) of wavelength
$\lambda=\frac{2} {n}=\frac{2\pi}{2k_{{\rm F}}}$ at zero
temperature.

The spin-spin correlation function $S(r)$ $=$ $\left\langle
n_{3}(r)\,n_{3} (0)\right\rangle$ can be obtained from
\begin{equation}
S(r)=\frac{n(2-n)}{n+2D-n^{2}}\sum_{a,b,c=1}^{2}I_{aa}^{-1}Q_{abac}(r)
\end{equation}
We can now establish the following relation between the spin and
charge correlation functions
\begin{equation}
S(r)=\frac{n-2D}{n+2D-n^{2}}\left[  N(r)-n^{2}\right]
\end{equation}
which implies they have the same spatial dependence. Consequently,
within the one-loop approximation, gaps in charge and/or spin
sectors will open simultaneously. This is also the situation in
the non-interacting case, where the spin-spin correlation function
reads as
\begin{equation}
S(r)=\delta_{r,0}n-\frac{1}{2}B^{2}(r)
\end{equation}
It is worth pointing out that this limitation is connected to the
use of the one-loop approximation and not to the Composite
Operator Method.

In the linear response theory, the spin susceptibility is defined
as $\chi _{s}(k,\omega)$$=-$${\cal F}\left\langle {\cal R}\left[
n_{3} (r,t)\,n_{3}(0)\right]  \right\rangle $ (${\cal F}$ is the
Fourier transform operator) and can be computed by means of the
following expression within the one-loop approximation
\begin{equation}
\chi_{s}(k,\omega)=-\frac{n(2-n)}{n+2D-n^{2}}\sum_{a,b,c=1}^{2}I_{aa}
^{-1}Q_{abac}^{R}(k,\omega)\label{spinsuseq}
\end{equation}
where
\begin{align}
Q_{abac}^{R}(k,\omega) &
=\frac{1}{2\pi}\sum_{m,n=1}^{2}\int_{-\pi}^{\pi
}dp\,\sigma_{ab}^{(n)}(k+p)\,\sigma_{ac}^{(m)}(p)\times\nonumber\\
&  \times\frac{f_{{\rm F}}\left[  E_{m}(p)\right] -f_{{\rm
F}}\left[ E_{n}(k+p)\right]
}{\omega+E_{n}(k+p)-E_{m}(p)+i\varepsilon}
\end{align}
The spin susceptibility, in the non-interacting case, is obtained
from the following integral
\begin{equation}
\chi_{s}(k,\omega)=-\frac{1}{\pi}\int_{-\pi}^{\pi}\frac{f_{{\rm
F}}\left[ E(p)\right]  -f_{{\rm F}}\left[  E(k+p)\right]  }{\omega
+E(p)-E(k+p)+i\varepsilon}dp
\end{equation}
At zero temperature, the uniform and static spin susceptibility
reads as
\begin{equation}\label{ChisU0}
\chi_{s}=\frac{1}{\pi\,t\sin\frac{\pi n}{2}}=2N_{cc}(0)
\end{equation}
The charge susceptibility is defined as $\chi_{c}(k,\omega)={\cal
F} \left\langle {\cal R}\left[ n(r,t)\,n(0)\right]  \right\rangle
$ and it can be computed as follows
\begin{equation}
\chi_{c}(k,\omega)=-\frac{n(2-n)}{n-2D}\sum_{a,b,c=1}^{2}I_{a}^{-1}
Q_{abac}^{R}(k,\omega)
\end{equation}
Once again, the spin and charge susceptibilities satisfy the
following relation
\begin{equation}
\chi_{s}(k,\omega)=\frac{n-2D}{n+2D-n^{2}}\chi_{c}(k,\omega)\label{susCH-SP}
\end{equation}
In the non-interacting case, the charge susceptibility coincides
with the spin one (i.e., $\chi_{c}(k,\omega)=\chi_{s}(k,\omega)$).

The uniform and static spin and charge susceptibilities can be
also calculated directly from their thermodynamical definitions
\begin{align}
\chi_{s}  &  =\left.  \frac{\partial m}{\partial h}\right|
_{h=0}\label{spinsusmh}\\
\chi_{c}  &  =\frac{\partial n}{\partial\mu}\label{chargesuseq}
\end{align}
where $m$ and $h$ are the magnetization per site and the external
applied field, respectively.

It is worth mentioning that all the expressions given in this
Section, computed within the Composite Operator Method, exactly
reproduces the non-interacting and the atomic limits.

\section{The Results}

\label{Results}

\begin{figure}[tb]
\begin{center}
\includegraphics[width=8cm]{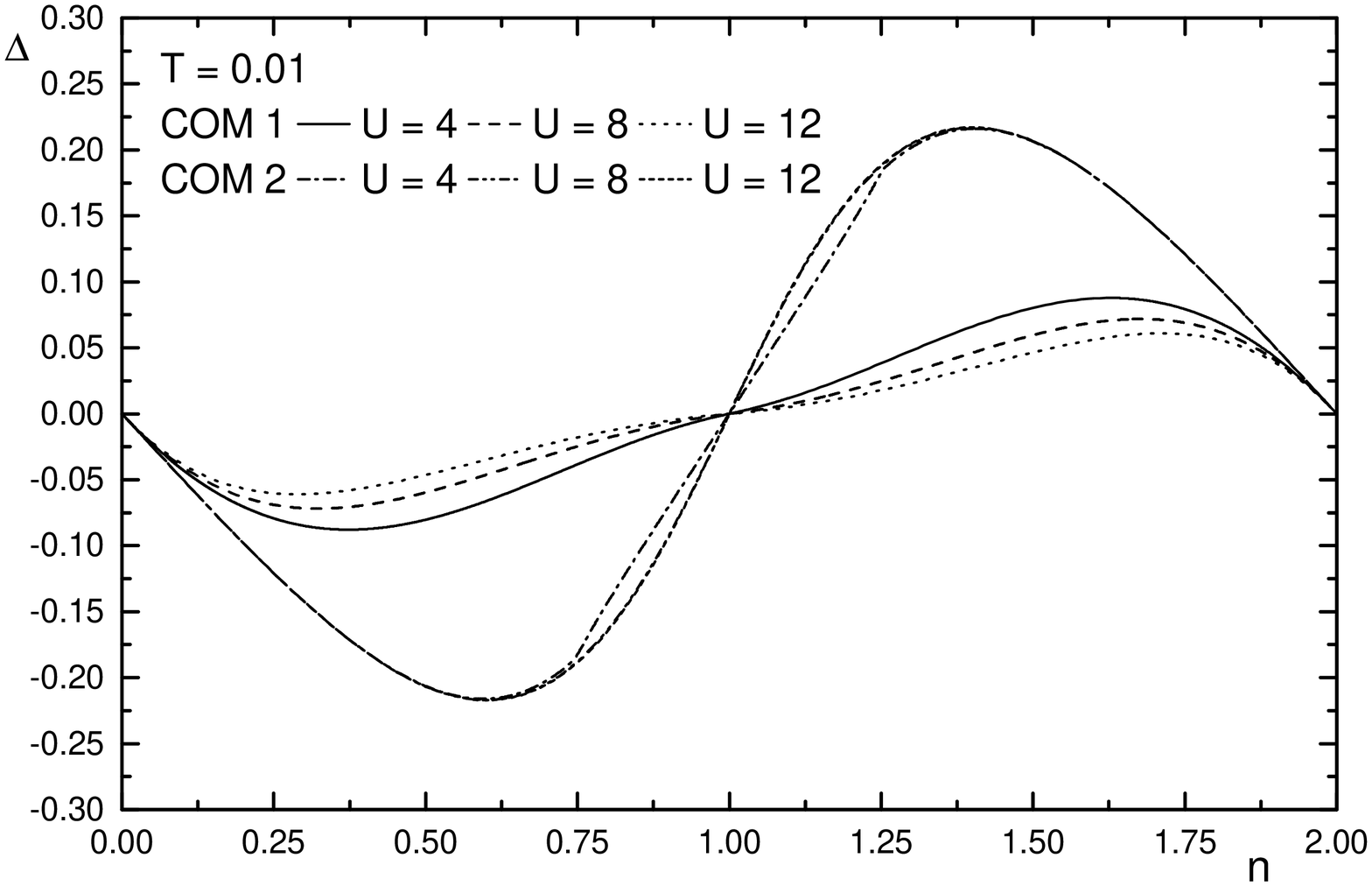}
\end{center}
\caption{\emph{Internal} parameter $\Delta$ as function of $n$ for
$T=0.01$ and $U=4$, $8$ and $12$.} \label{Fig1}
\end{figure}

\begin{figure}[tb]
\begin{center}
\includegraphics[width=8cm]{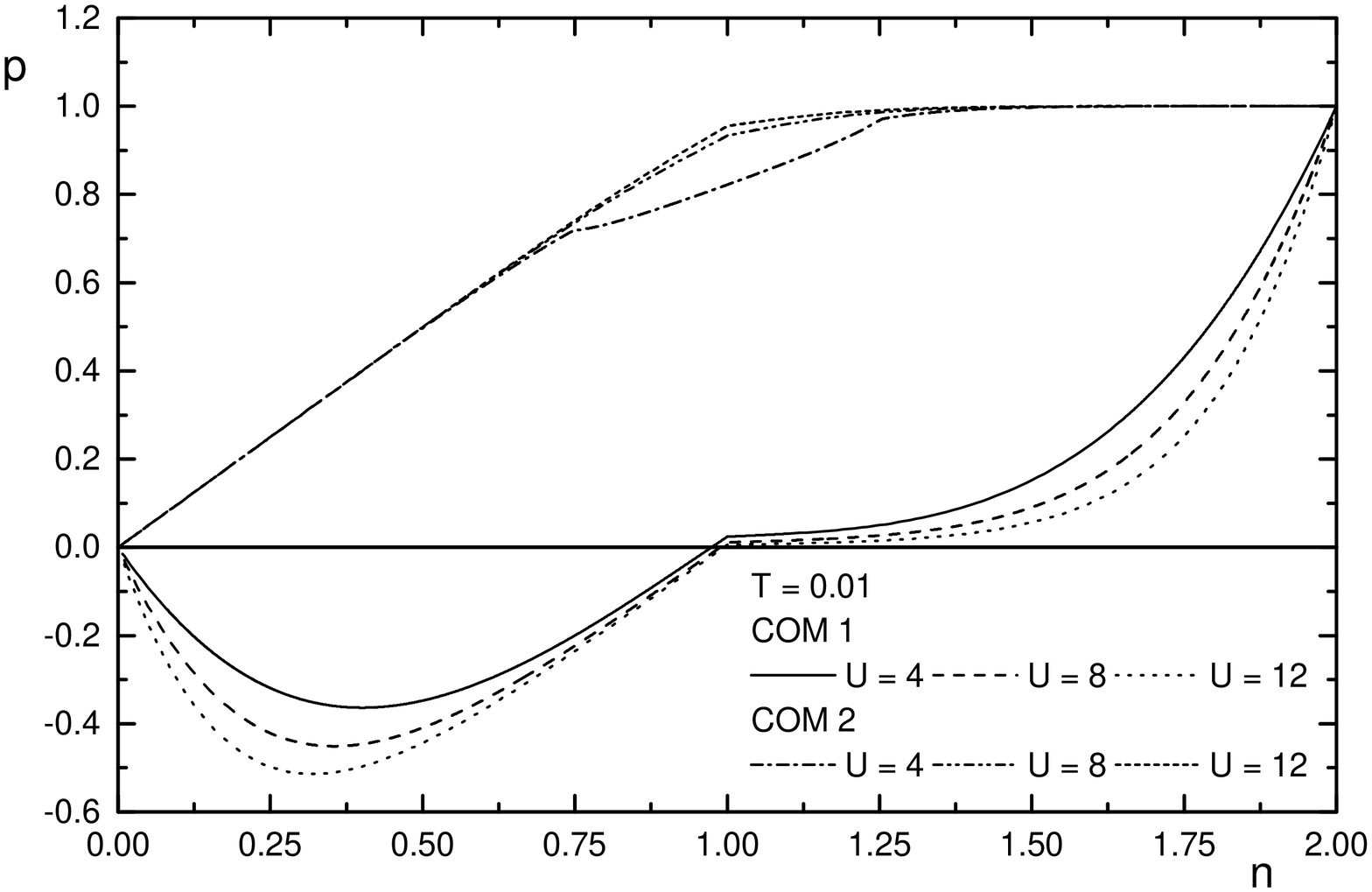}
\end{center}
\caption{\emph{Internal} parameter $p$ as function of $n$ for
$T=0.01$ and $U=4$, $8$ and $12$.} \label{Fig3}
\end{figure}

\begin{figure}[tb]
\begin{center}
\includegraphics[width=8cm]{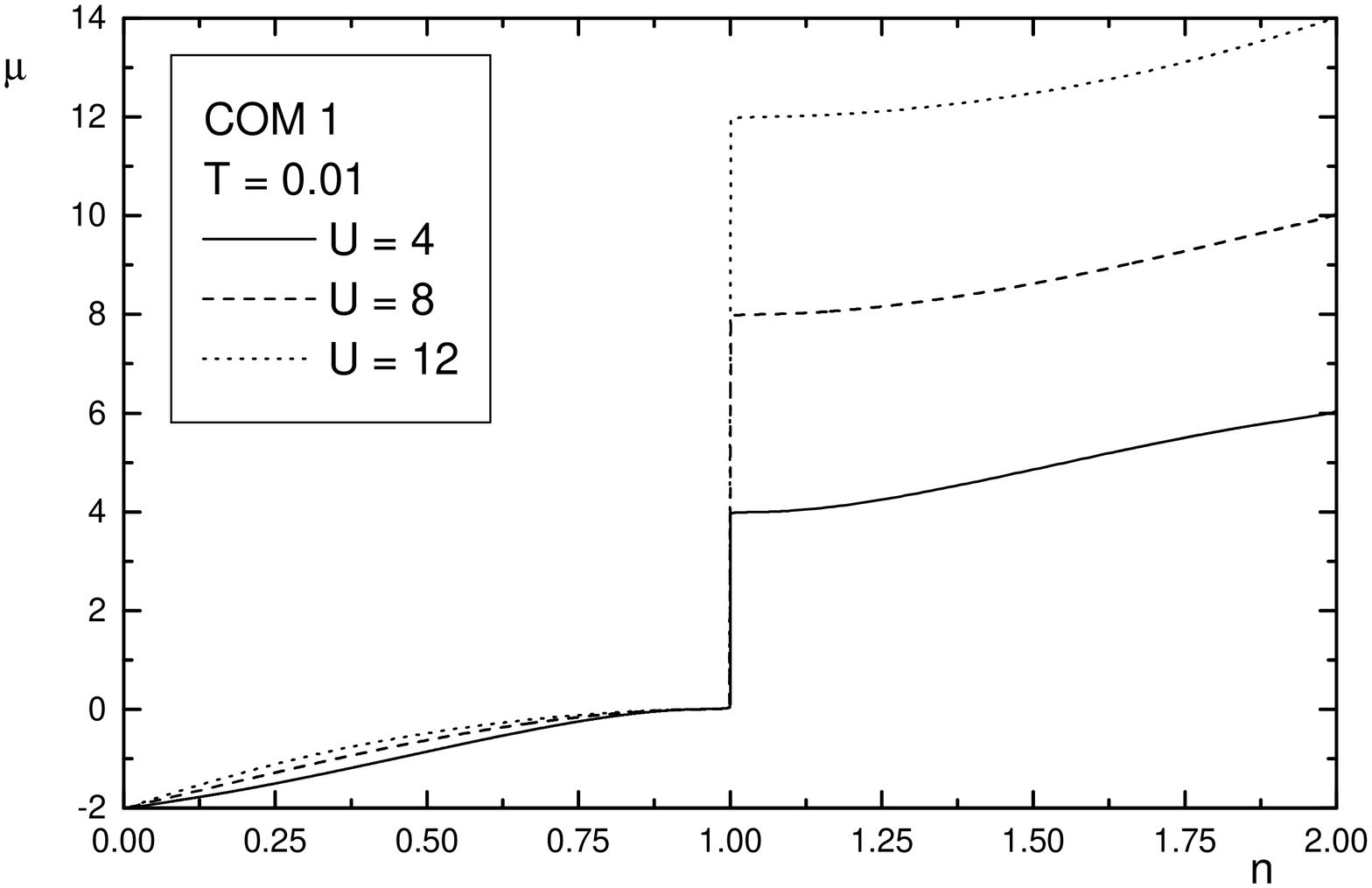}
\end{center}
\caption{Chemical potential $\mu$ as function of $n$ for $T=0.01$
and $U=4$, $8$ and $12$ (\emph{COM}~1 solution).} \label{Fig5}
\end{figure}

\begin{figure}[tb]
\begin{center}
\includegraphics[width=8cm]{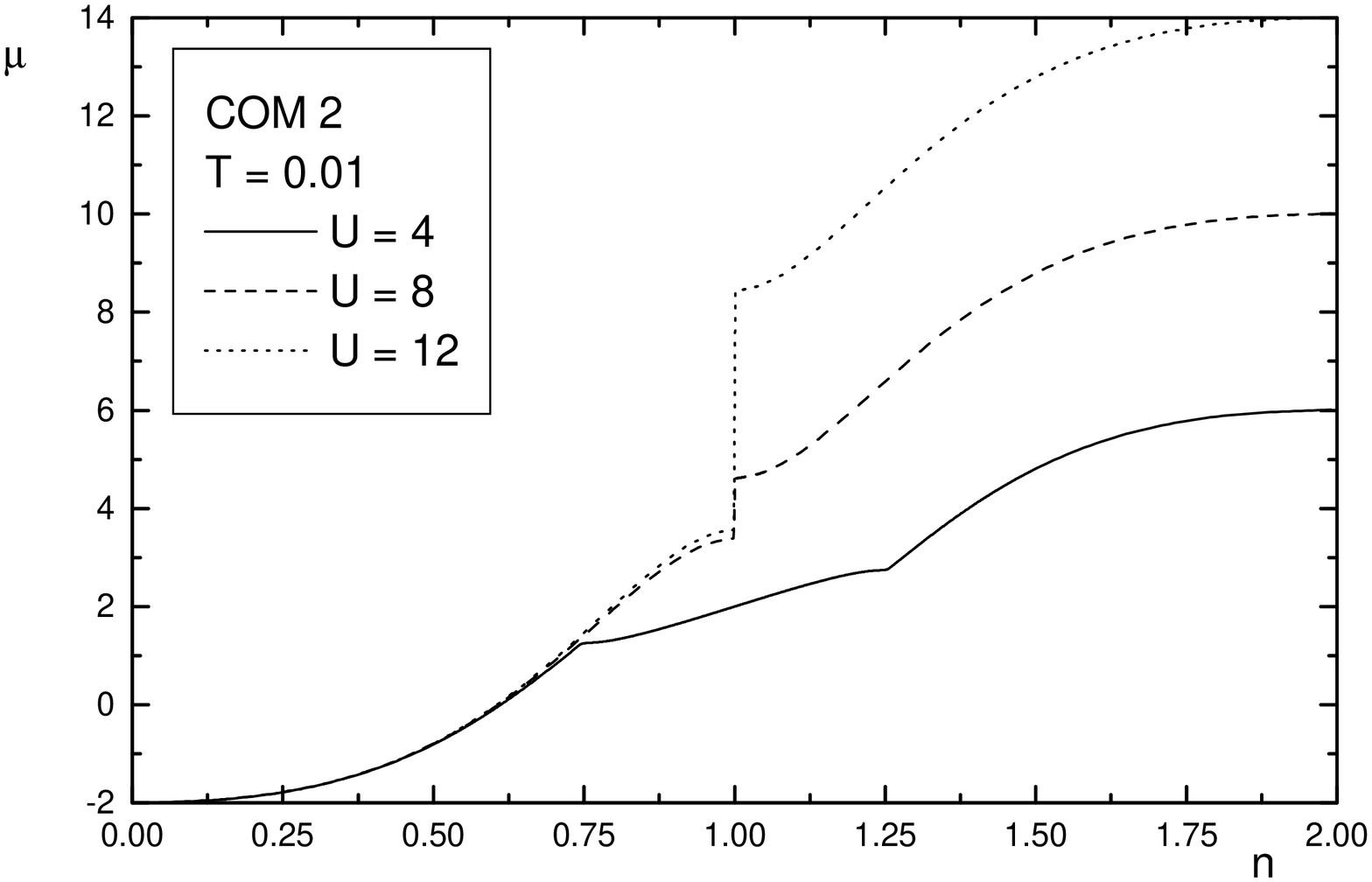}
\end{center}
\caption{Chemical potential $\mu$ as function of $n$ for $T=0.01$
and $U=4$, $8$ and $12$ (\emph{COM}~2 solution).} \label{Fig6}
\end{figure}

\begin{figure}[tb]
\begin{center}
\includegraphics[width=8cm]{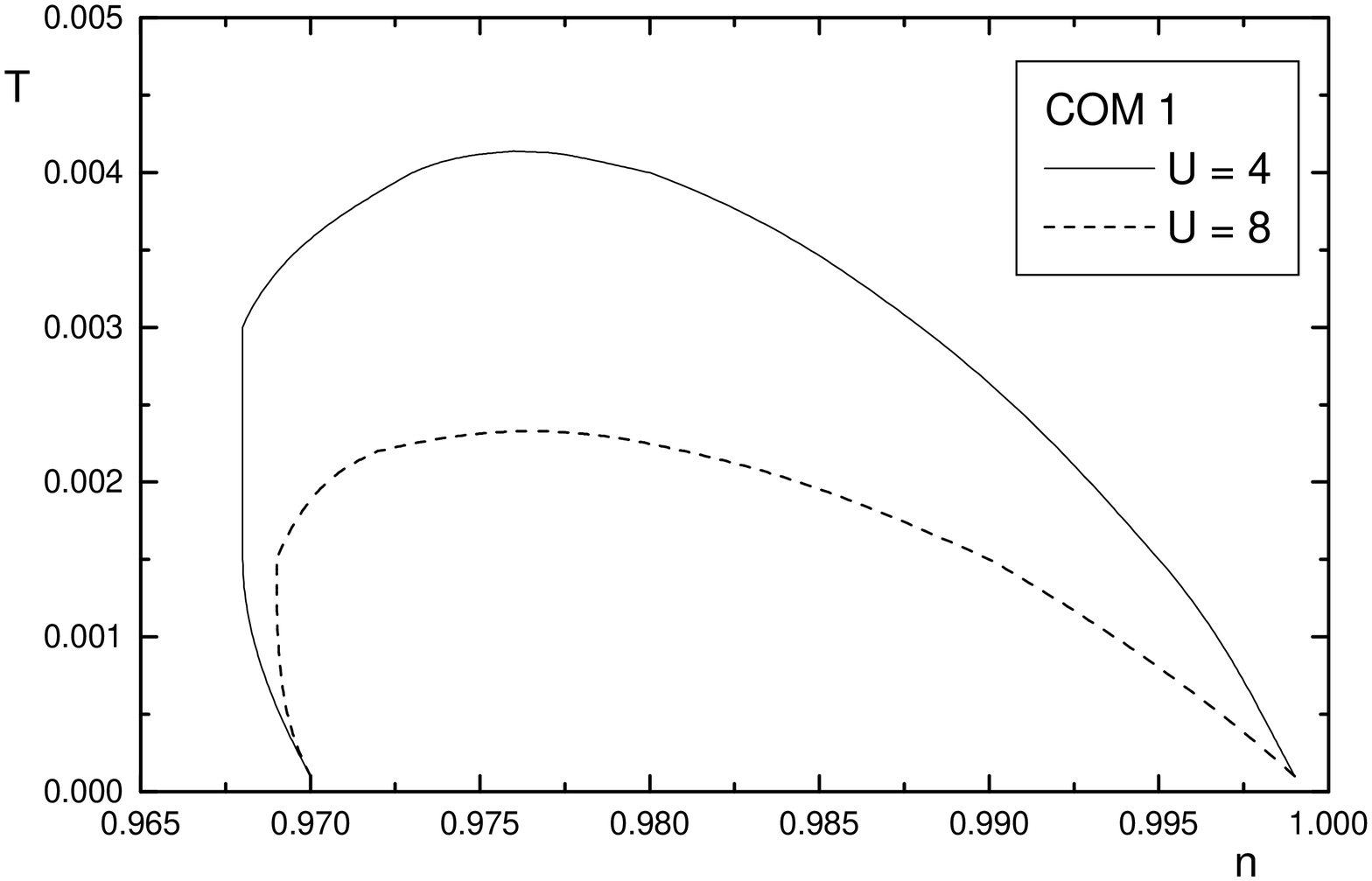}
\end{center}
\caption{Instability region within the plane $n$-$T$ for $U=4$ and
$8$ (\emph{COM}~1 solution).} \label{Fig7}
\end{figure}

The set of self-consistent equations in (\ref{self}), that
determines the \emph{internal} parameters, admits two distinct
solutions. Hereafter, we will call these solutions \emph{COM}~1
and \emph{COM}~2. The evolution of the \emph{internal} parameters
with the \emph{external} ones (i.e., filling, intrasite Coulomb
potential and temperature) reveals substantial differences between
the two solutions. The $\Delta$ parameter is smaller and much more
sensitive to the strength of Coulomb interaction in \emph{COM}~1
(see Fig.~\ref{Fig1}). For $n\leq1$, the $p$ parameter is negative
or very small and positive in \emph{COM}~1, while it is always
positive and of the order of the filling in \emph{COM}~2 see
Fig.~\ref{Fig3}; actually, at half filling and on increasing $U$,
$p$ tends to $0$ in \emph{COM}~1 and to $1$ in \emph{COM}~2. In
\emph{COM}~1, the chemical potential $\mu$ shows a discontinuity
at half filling for any finite value of the Coulomb interaction
and a zone of instability (i.e., a negative compressibility) at
small doping, temperatures and interaction strength (see
Fig.~\ref{Fig5}). In \emph{COM}~2, the discontinuity of the
chemical potential appears only after a critical value of the
Coulomb interaction is reached (see Fig.~\ref{Fig6}). As we will
show in the next section, the absence of the Mott-Hubbard
transition, which is a consequence of the mainly negative value of
the $p$ parameter, plays a key role in the physics described by
the \emph{COM}~1 solution. The instability of \emph{COM} ~1 can be
studied by looking at the compressibility (i.e.,
$\kappa=\frac1{n^{2}}\frac{\partial n}{\partial\mu}$). The
instability is confined to a very small region in the plane
$(n,T)$. This region does not comprehend half filling (see
Fig.~\ref{Fig7}). For the \emph{2D} Hubbard model, which also has
two solutions within the framework of the \emph{COM}, the region
of instability is much larger.

\subsection{Properties at half filling}

\begin{figure}[tb]
\begin{center}
\includegraphics[width=8cm]{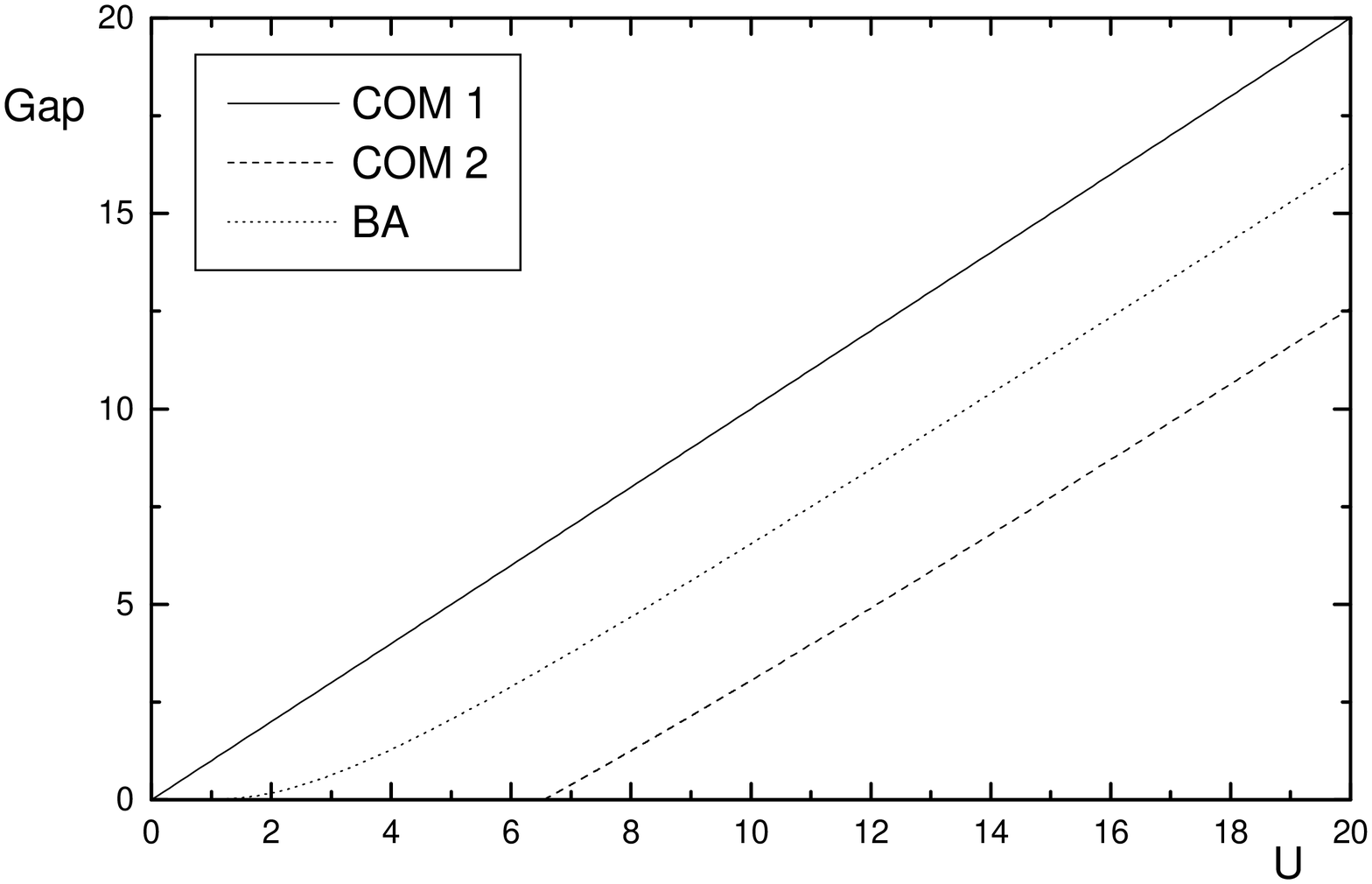}
\end{center}
\caption{Gap as function of $U$ (\emph{COM}~1 solution).}
\label{Fig8}
\end{figure}

\begin{figure}[tb]
\begin{center}
\includegraphics[width=8cm]{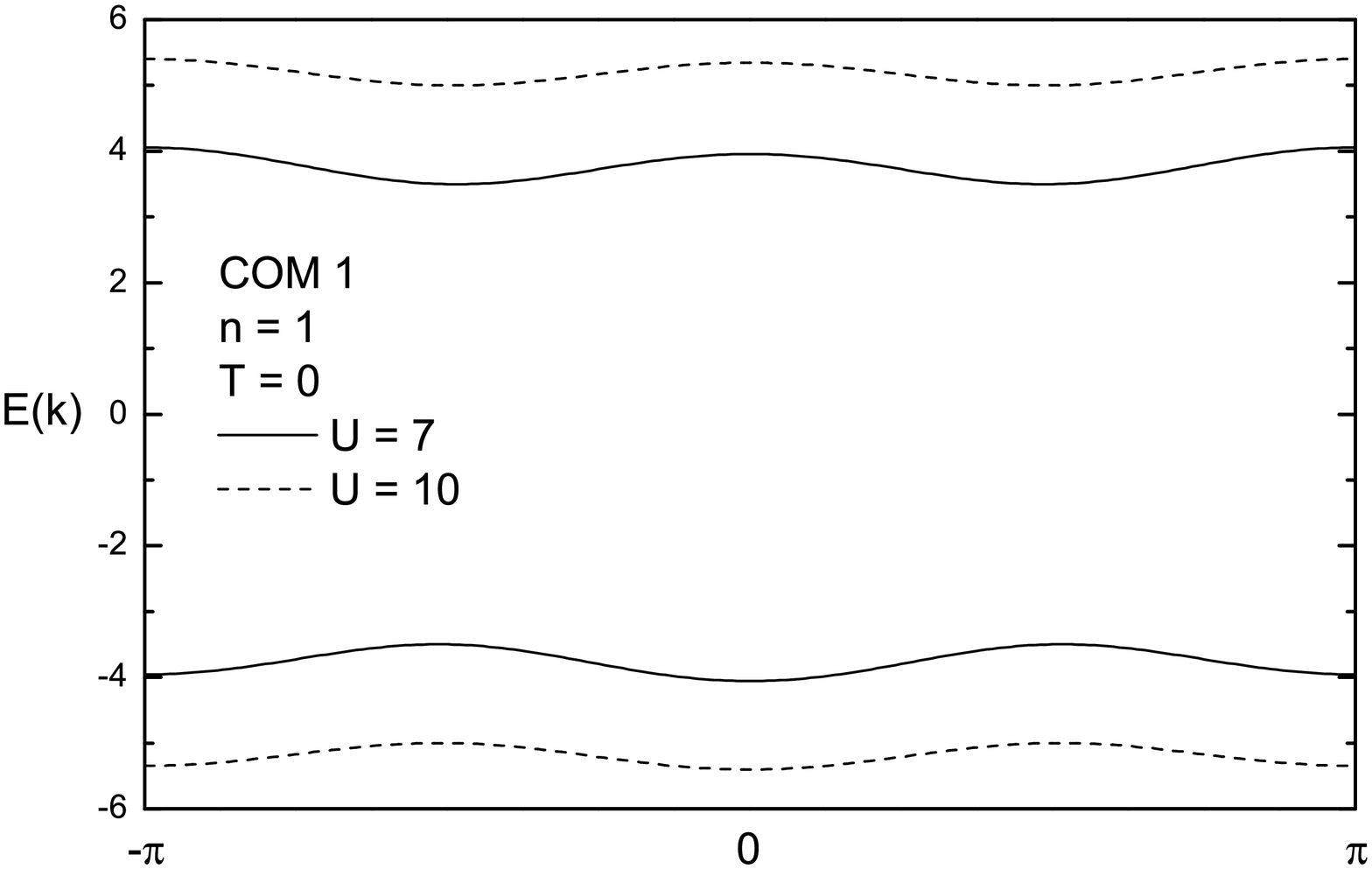}
\end{center}
\caption{Energy spectra $E(k)$ for $T=0$, $n=1$ and $U=7$ and $10$
(\emph{COM}~1 solution).} \label{Fig9}
\end{figure}

\begin{figure}[tb]
\begin{center}
\includegraphics[width=8cm]{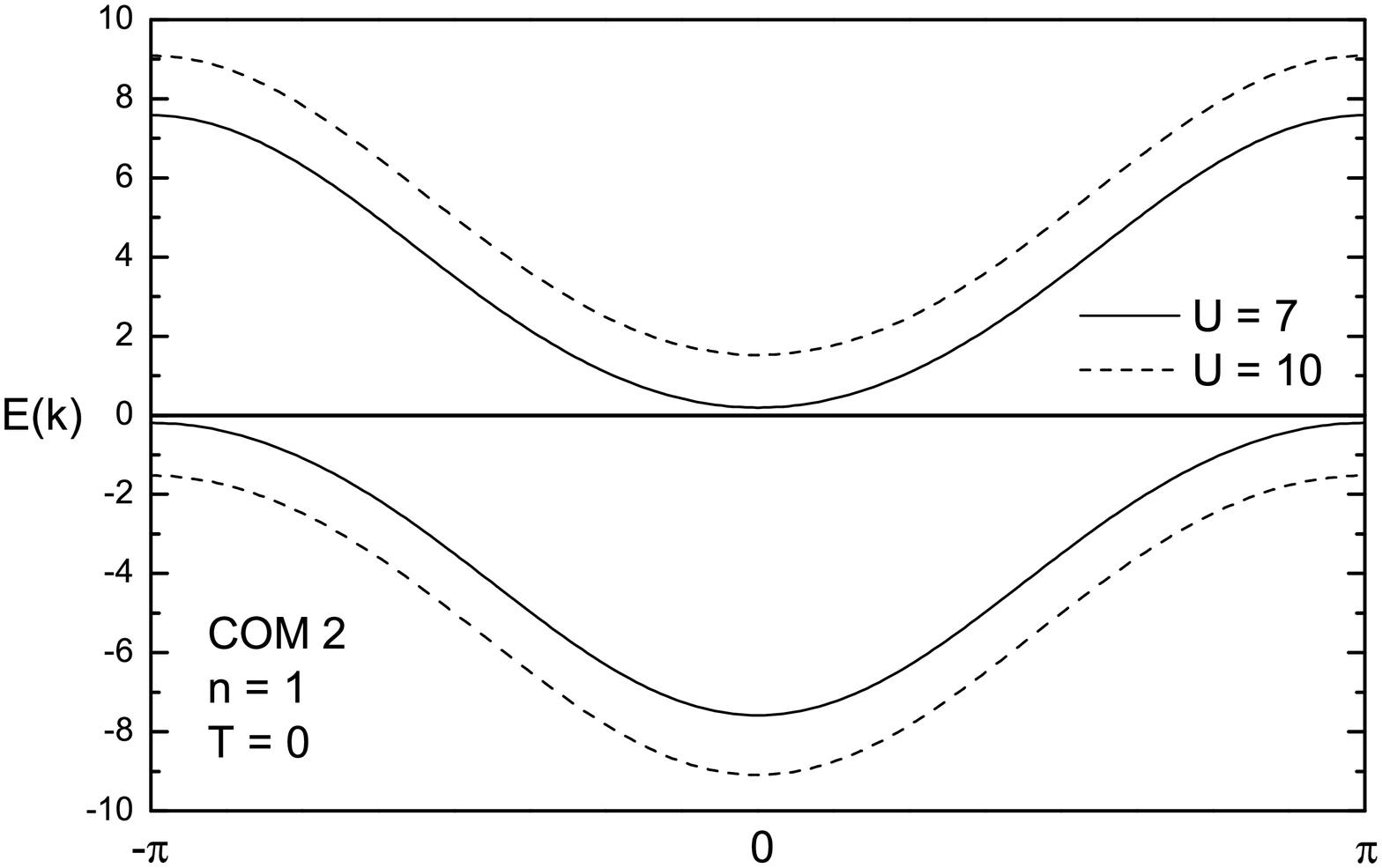}
\end{center}
\caption{Energy spectra $E(k)$ for $T=0$, $n=1$ and $U=7$ and $10$
(\emph{COM}~2 solution).} \label{Fig10}
\end{figure}

\begin{figure}[tb]
\begin{center}
\includegraphics[width=8cm]{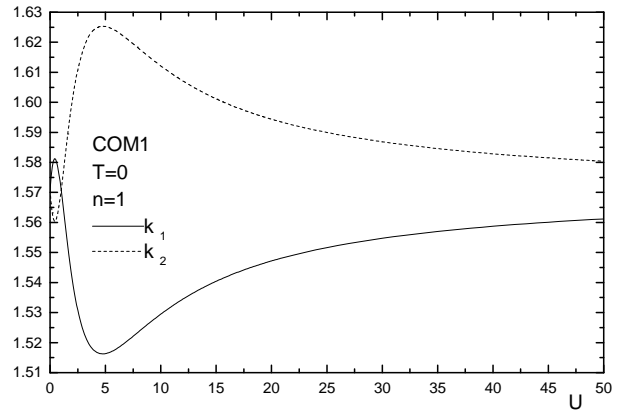}
\end{center}
\caption{$k_1$ and $k_2$, see definition in the text, as function
of $U$ for $T=0$ and $n=1$ (\emph{COM}~1 solution).} \label{Fig11}
\end{figure}

The ground-state properties of the \emph{1D} Hubbard model at half
filling were exactly derived by Lieb and Wu using the
\emph{BA}\cite{Lieb:68}. This exact solution corresponds to an
insulating state with short-range antiferromagnetic (\emph{AF})
correlations for any finite value of $U$. According to this, we
will compare the evolution of the gap in the excitation spectrum
and the band structure of both solutions with the \emph{BA } exact
results, in order to choose the solution that gives the most
consistent physical picture at half filling.

As already mentioned above, within the analysis of the chemical
potential, the \emph{COM}~1 solution presents a gap for any finite
value of the Coulomb interaction, in agreement with the \emph{BA}
result, while the \emph{COM}~2 solution is characterized by a
critical value of the Coulomb interaction (i.e.,
$U_{c}\approx6.56$) above which a gap opens (see Fig.~\ref{Fig8}).
The rate at which the gap opens in \emph{COM}~1 coincides with
that of the exact solution for $U\geq4$. The band structure (i.e.,
the excitation spectrum) of the two solutions is another
interesting property to be studied and compared. In fact, many
features, more or less anomalous, of both solutions can be easily
understood just looking at their spectra. As it can be seen in
Fig.~\ref{Fig9}, the \emph{COM}~1 solution has a typical \emph{AF
} band pattern (i.e., a quasi-halved Brillouin zone, the first
excitation around $k=\pm\frac{\pi}{2}$ and a very narrow bandwidth
of the order $J= \frac {4t^{2}}{U}$), in agreement with the
\emph{BA} result. On the contrary, the band structure of the
\emph{COM}~2 solution corresponds to a typically paramagnetic
state, with the first hole excitation at $k=\pm\pi$, the first
electron excitation at $k=0$ and a bandwidth of the order $8t$
(see Fig.~\ref{Fig10}). In the figures, the energy is measured
with respect to the chemical potential. In particular, while
\emph{COM}~2 has two subbands with both a minimum in $k=0$ and a
maximum in $k=\pm\pi$, the \emph{COM}~1 solution has the upper
subband with a maximum in $k=\pm\pi$ and a minimum at $k_{1}
=\pm\arccos\theta_{0}$ and the lower subband with a maximum at
$k_{2} =\pi-k_{1}$ and a minimum in $k=0$, where
\begin{equation}
\theta_{0}=\frac{p\,U}{2t(1-2p)\sqrt{1-4p}}
\end{equation}
For large values of the Coulomb interaction both $k_{1}$ and
$k_{2}$ tend to $\frac{\pi}{2}$ since $p$ tends to zero (see
Fig.~\ref{Fig11}). According to this analysis, the width of the
subbands and the value of the gap in both solutions can be easily
computed. In \emph{COM}~2, the width of the subbands at half
filling is $W=8t\,p$, which tends to $8t$ for large values of the
Coulomb interaction, while the gap, above the critical value
$U_{c} =4t\sqrt{4p-1}\approx6.56$, has the expression
\begin{equation}
\Delta E=-8t\,p+\sqrt{U^{2}+16t^{2}(2p-1)^{2}}\label{DelE2}
\end{equation}
On the contrary, in \emph{COM}~1 the width of the subbands at half
filling is
\begin{equation}
W=4t\,p+\frac{1}{2}\sqrt{U^{2}+16t^{2}(2p-1)^{2}}-U\frac{\sqrt{1-4p}}{2(1-2p)
}
\end{equation}
which tends to $J=\frac{4t^{2}}{U}$ for large values of the
Coulomb interaction, while the gap, has the expression
\begin{equation}
\Delta E=\frac{\sqrt{1-4p}}{1-2p}U\label{DelE1}
\end{equation}
Both expressions (i.e., Eqs.~(\ref{DelE2}) and (\ref{DelE1})) for
the gap tend to $U$ for large values of the Coulomb interaction.
It is worth pointing out that, within the two-pole approximation,
the $p$ parameter rules the opening of the gap at half filling. In
particular, if $p<\frac{1}{4}\left( 1+\frac{U^{2}}{16t^{2}}\right)
$ we have a gapped solution, because the two subbands have
opposite signs for any value of momentum and therefore they do not
overlap. This is the case for the \emph{COM}~1 solution for any
value of the Coulomb interaction. For the \emph{COM}~2 solution we
have a gap above the critical value $U_{c}$. Otherwise, we have no
gap and the two subbands overlap. In particular, both subbands
have negative values for $|k|<\arccos x_{0}$, where
$x_{0}=U/U_{c}$.

We can conclude that only \emph{COM}~1 gives a description of the
physics of the half-filled \emph{1D} Hubbard model consistent with
the exact results obtained by the \emph{BA}. Therefore, in this
section, we will mainly focus on this solution.

\subsubsection{Local properties}

\begin{figure}[tb]
\begin{center}
\includegraphics[width=8cm]{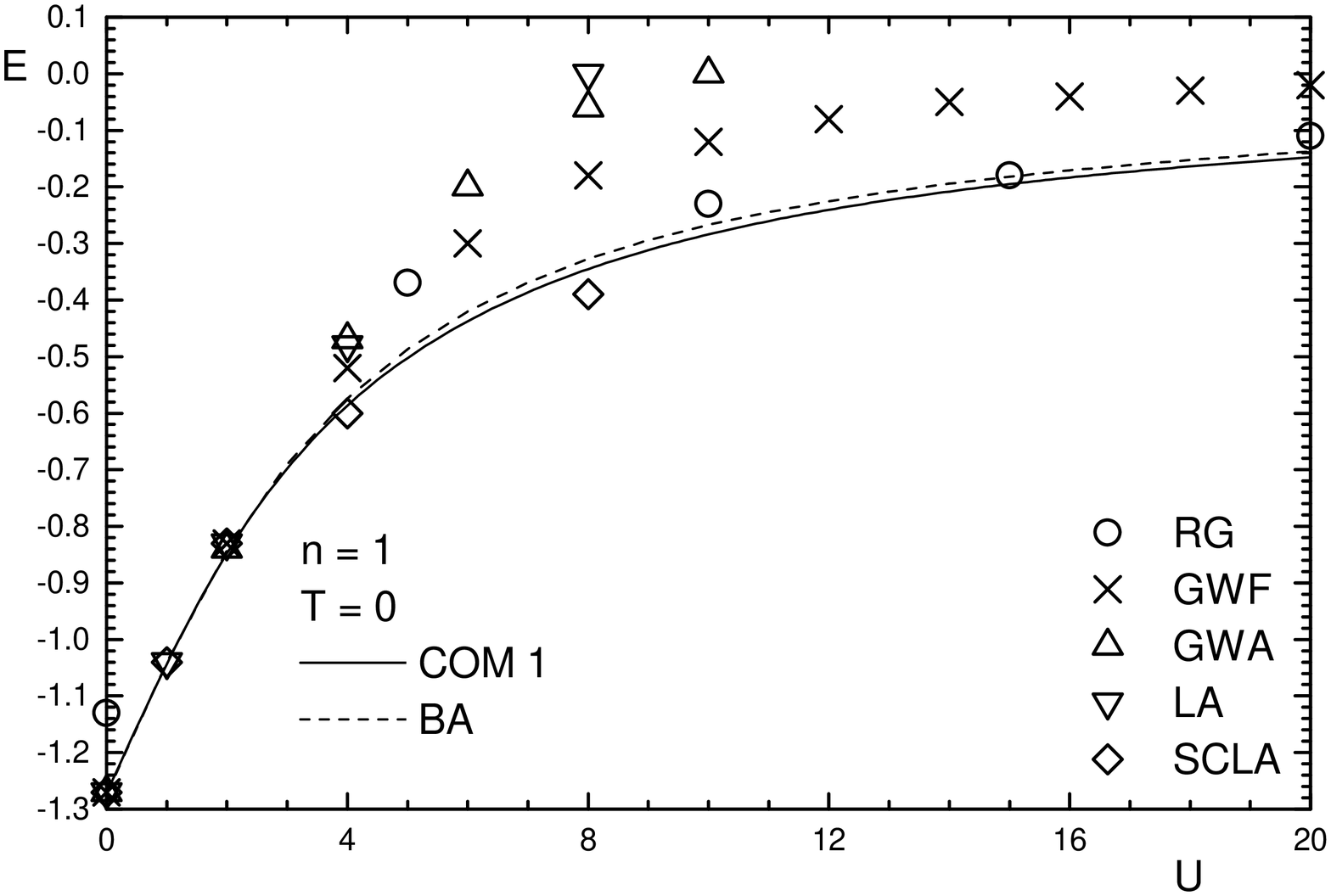}
\end{center}
\caption{Internal energy $E$ as function of $U$ for $T=0$ and
$n=1$ (\emph{COM}~1 solution).} \label{Fig12}
\end{figure}

\begin{figure}[tb]
\begin{center}
\includegraphics[width=8cm]{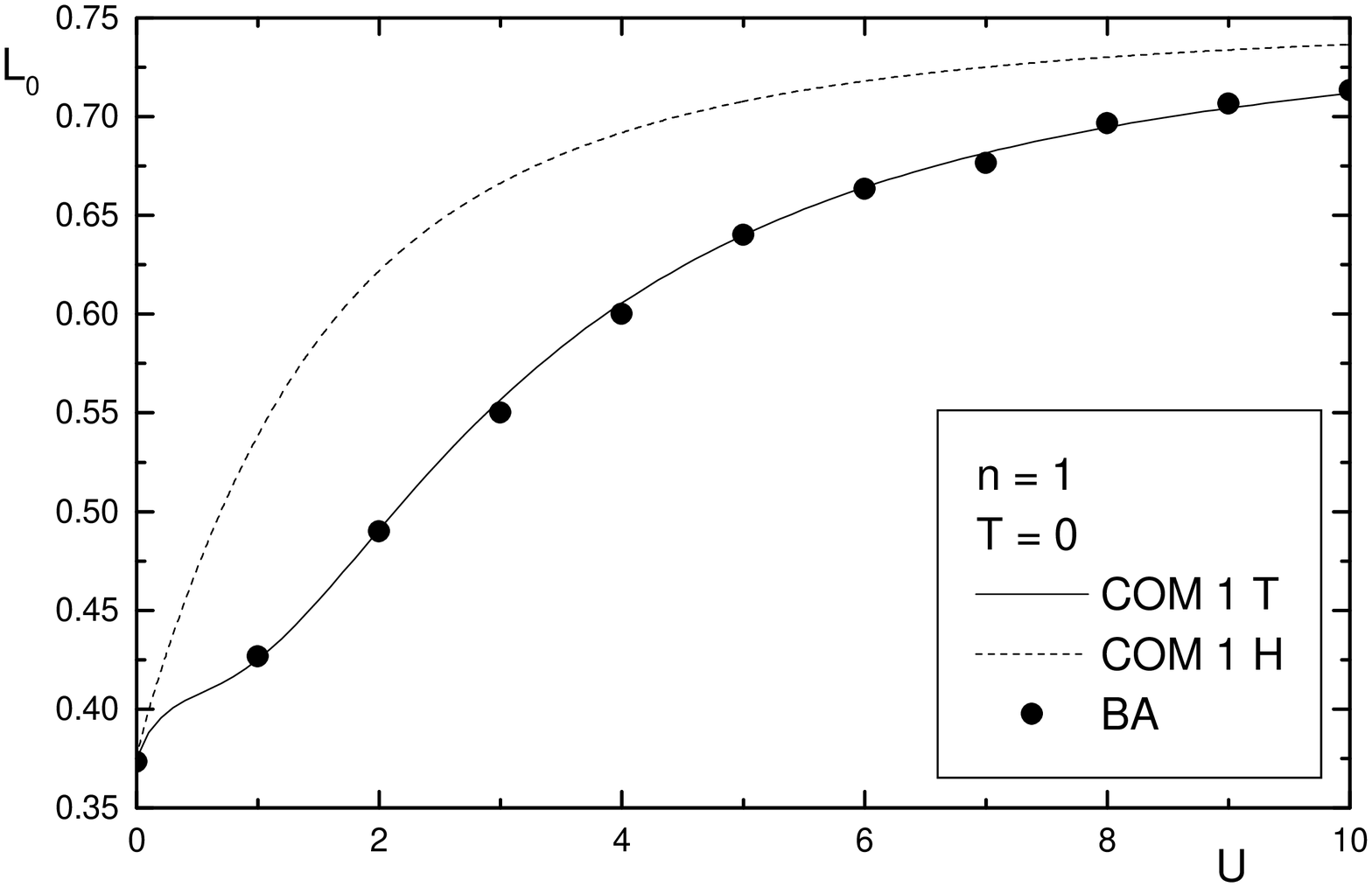}
\end{center}
\caption{Local magnetization $L_{0}$ as function of $U$ for $T=0$
and $n=1$.} \label{Fig13}
\end{figure}

The internal energy, at half filling and zero temperature, can be
exactly calculated by means of the \emph{BA} through
Eq.~(\ref{EBA}). The two relevant limits (i.e., small and large
Coulomb interaction) read as follows
\begin{align}
\lim_{U\rightarrow0}E= &
-4\frac{|t|}{\pi}+\frac{1}{4}U+O(\frac{U^{2}}{t})\nonumber\\
\lim_{U\rightarrow\infty}E= &
-4\frac{t^{2}}{U}\ln2+O(\frac{t^{4}}{U^{3}})
\end{align}
where $4\ln2\cong2.77$. The \emph{COM} solution exactly agrees
with the \emph{BA} result in the weak-interacting limit, while in
the strong-interacting limit, we have
\begin{align}
\lim_{U\rightarrow\infty}E_{H}^{\emph{COM}1}= &
-3\frac{t^{2}}{U}+O(\frac{t^{4}}{U^{3}})\nonumber\\
\lim_{U\rightarrow\infty}E_{H}^{\emph{COM}2}= &
5\frac{t^{2}}{U}+O(\frac{t^{4}}{U^{3}})
\end{align}
Again, while \emph{COM}~1 gives a result very close to the
\emph{BA} one, the \emph{COM}~2 solution is very far in this
limit. The internal energy $E$ at half filling and zero
temperature, calculated by means of Eq.~(\ref{EH}) in the
\emph{COM}~1 solution (i.e., $E_{H}^{\emph{COM}1}$), is shown, as
a function of the Coulomb interaction strength, in
Fig.~\ref{Fig12}. The results obtained by means of the \emph{BA}
\cite{Shiba:72} and other analytical
approaches\cite{Hirsch:80,Metzner:87,Buzatu:95} are also reported.
As we can see, the agreement between \emph{COM}~1 and \emph{BA} is
excellent. The self-consistent Ladder approximation (\emph{SCLA})
of Ref.~\onlinecite {Buzatu:95} shows also a very good agreement
for all values of the coupling, but it does not have the correct
behaviour for an infinite value of the Coulomb interaction.
Moreover, both the Ladder (\emph{LA}) \cite{Buzatu:95} and the
Gutzwiller (\emph{GWA} and \emph{GWF}) \cite{Metzner:87}
approximations go to zero at finite $U$, whereas the
Renormalization Group (\emph{RG})\cite{Hirsch:80} has the right
asymptotic behaviour for $U\mapsto\infty$, but it does not
reproduce the non-interacting limit. The local magnetization
$L_{0}$ at half filling and zero temperature as a function of the
interaction strength is shown in Fig.~\ref{Fig13}. We report both
the \emph{COM} results (i.e., $L_{0H}^{\emph{COM}1}$ and
$L_{0T}^{\emph{COM}1}$) and the results obtained by means of the
\emph{BA}\cite{Shiba:72a}. $L_{0T}^{\emph{COM}1}$ is in excellent
agreement with the exact solution. As one should expect, the
electron localization increases with $U$ and for infinite $U$
reaches a saturation (i.e., zero double occupancy and zero kinetic
energy). Thus, the \emph{1D} itinerant electron system described
by the Hubbard chain is equivalent, at half filling and infinite
$U$, to the system of localized spins described by the
spin-$\frac{1}{2}$ \emph{AF} Heisenberg model.

\subsubsection{Thermodynamic properties}

\begin{figure}[tb]
\begin{center}
\includegraphics[width=8cm]{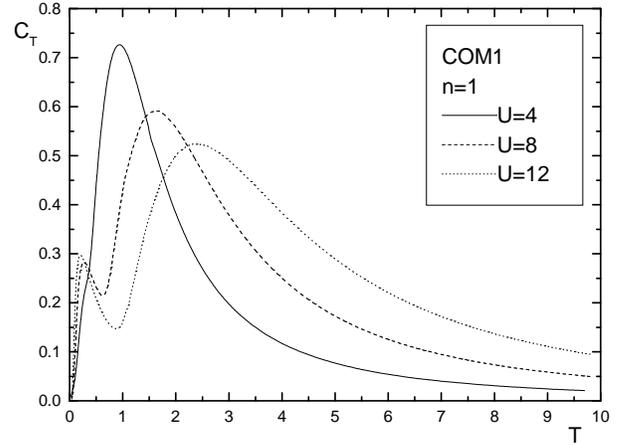}
\end{center}
\caption{Specific heat $C_{T}$ as function of $T$ for $n=1$ and
$U=4$, $8$ and $12$ (\emph{COM}~1 solution).} \label{Fig14}
\end{figure}

\begin{figure}[tb]
\begin{center}
\includegraphics[width=8cm]{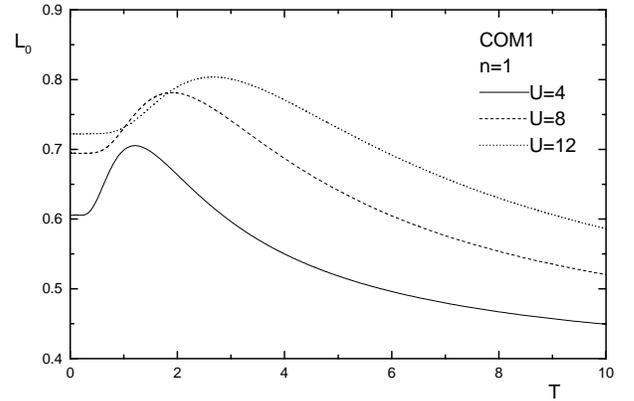}
\end{center}
\caption{Local magnetization $L_{0}$ as function of $T$ for $n=1$
and $U=4$, $8$ and $12$ (\emph{COM}~1 solution).} \label{Fig15}
\end{figure}

\begin{figure}[tb]
\begin{center}
\includegraphics[width=8cm]{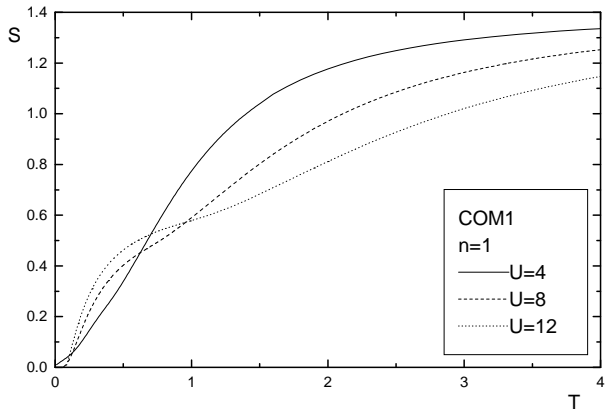}
\end{center}
\caption{Entropy $S$ as function of $T$ for $n=1$ and $U=4$, $8$
and $12$ (\emph{COM}~1 solution).} \label{Fig16}
\end{figure}

The thermodynamic properties of the Hubbard chain can be evaluated
using the \emph{BA} by means of the finite-temperature formalism
developed by Takahashi \cite{Takahashi:72}.

The specific heat $C_{T}$ at half filling, calculated by means of
Eq.~\ref{cv} in the \emph{COM}~1 solution (i.e.,
$C_{T}^{\emph{COM}1}$), is shown in Fig.~\ref{Fig14} as a function
of the temperature for different values of the Coulomb
interaction. By increasing the interaction strength, the single
peak present for $U\leq W=4t$ splits in two peaks moving to
opposite directions in temperature ($W$ is the non-interacting
bandwidth). The low-$T$ feature is due to spin excitations: it is
located at $T\sim J$, where $J=\frac{4t^{2}}{U}$ is the magnitude
of the induced \emph{AF} exchange parameter. Such a low-$T$ peak
is characteristic of the \emph{AF} Heisenberg
chain\cite{Bonner:64}. This feature is consistent with the
\emph{AF}-like band structure of the \emph{COM}~1 solution (cfr.
Fig.~\ref{Fig9}). The high-$T$ peak is obviously associated to
charge excitations: it moves towards higher temperature as the gap
increases with $U$. As $T$ increases, the thermal energy allows
the electrons to be excited across the gap. Such a structure of
the specific heat, with low- and high- $T$ regions, dominated by
spin and charge excitations, respectively, is consistent with the
physical picture described by the exact \emph{BA}
solution\cite{Shiba:72a,Kawakami:89,Usuki:90,Juttner:98,Deguchi:00}
and it also agrees with recent \emph{qMC} calculations on finite
chains\cite{Schulte:96}.

In order to get information about the degree of localization of
the electrons we have also computed the local magnetization as a
function of the temperature. The results are reported in
Fig.~\ref{Fig15}. Let us note that the maximum localization occurs
at finite temperature. This is due to the strong antiferromagnetic
correlations present at zero temperature. These correlations
require a virtual hopping and, therefore, diminish the degree of
localization. Only by increasing the temperature we can suppress
the antiferromagnetic correlations and increase the localization.
Note also that the flat regions at low temperatures end at the
same temperature where the local minima develop in $C_{T}$. This
further confirms the spin-nature of the low-$T$ peak in the
specific heat.

In the temperature evolution of the entropy at half filling, spin
and charge degrees of freedom manifest separately in the
strong-interacting regime $U\geq W$ (see Fig.~\ref{Fig16} where
$S_{T}^{\emph{COM}1}$ is calculated by means of Eq.~\ref{ST} in
the \emph{COM}~1 solution). The low-$T$ region is dominated by
spin excitations and the high-$T$ region by particle excitations.
The internal energy shows an analogous behavior. The borderline
between these two regions can be set at $T\sim t$ in very good
agreement with both \emph{BA} \cite{Kawakami:89,Usuki:90} and
numerical\cite{Shiba:72a} results. It is worth noting that, in our
solution, the entropy shows the correct limiting behavior for high
temperatures (i.e.,
$\lim_{T\rightarrow\infty}S_{T}=2\ln2\cong1.39$).

\subsubsection{Correlation functions and susceptibilities}

\begin{figure}[tb]
\begin{center}
\includegraphics[width=8cm]{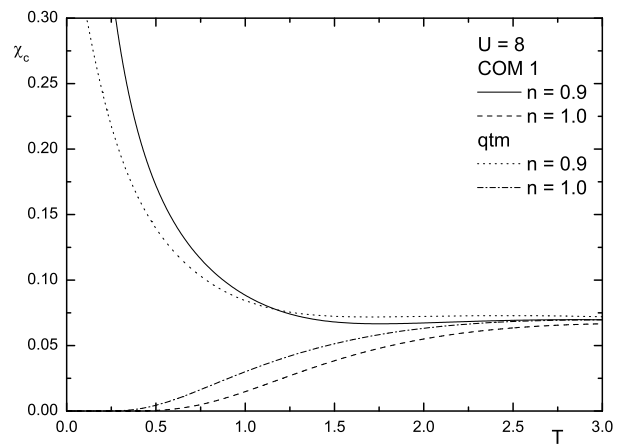}
\end{center}
\caption{Static and uniform charge susceptibility $\chi_{c}$ as
function of $T$ (\emph{COM}~1 solution).} \label{Fig17}
\end{figure}

\begin{figure}[tb]
\begin{center}
\includegraphics[width=8cm]{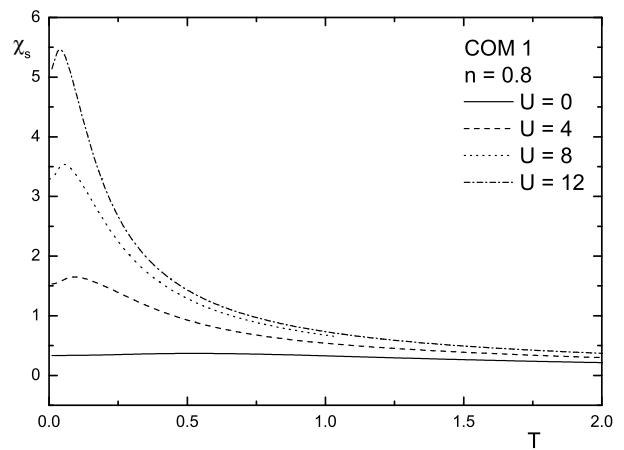}
\end{center}
\caption{Static and uniform spin susceptibility $\chi_{s}$ as
function of $T$ (\emph{COM}~1 solution).} \label{Fig18}
\end{figure}

\begin{figure}[tb]
\begin{center}
\includegraphics[width=8cm]{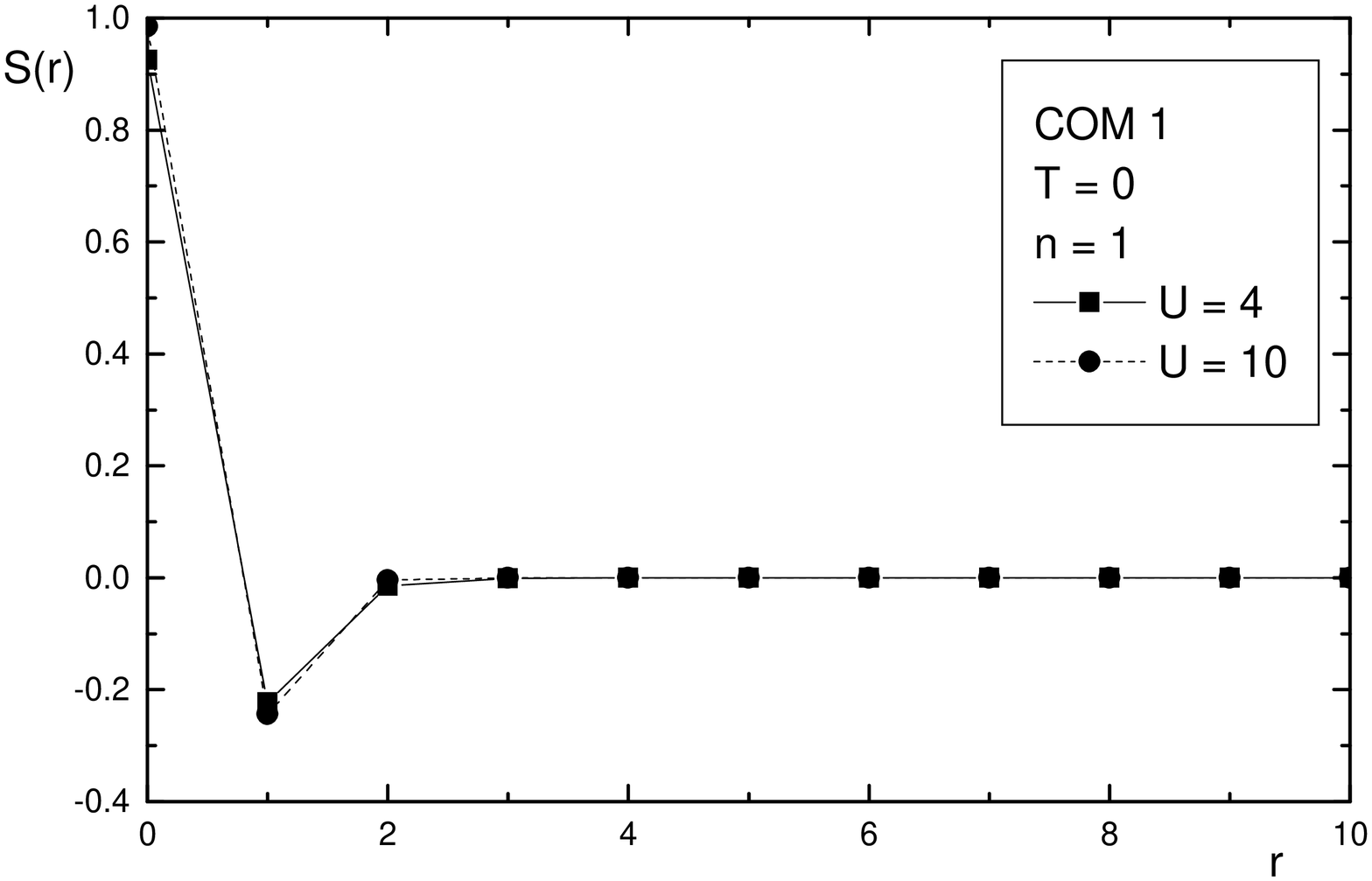}
\end{center}
\caption{Spin correlation function $S(r)$ for $T=0$, $n=1$ and
$U=4$, $10$ (\emph{COM}~1 solution).} \label{Fig19}
\end{figure}

A quantity which gives information about the evolution of the
charge excitations in the system is the charge susceptibility
($\chi_{c}$), which we calculate by means of Eq.
(\ref{chargesuseq}). This property is shown in Fig.~\ref{Fig17}
for $n=0.9$ and $n=1$; string theory results are taken from
Ref.~\cite{Deguchi:00}. The first thing to note is the strong
dependence of $\chi_{c}$ on the particle concentration at low
temperatures and for any coupling regime (not shown): the charge
susceptibility is strongly enhanced as $n$ approaches half
filling, while it goes to zero at $n=1$ as a consequence of the
opening of the gap in the charge excitation spectrum. It is worth
noticing our good agreement with the practically exact results of
\emph{qtm}\cite{Juttner:98}. For increasing Coulomb interaction
(not shown), the low-temperature enhancement of $\chi_{c}$ in the
low doping region is more evident, indicating that the charge
excitations are strongly renormalized by the Coulomb interaction
near the metal-insulator transition. At higher temperatures,
however, $\chi_{c}$ decreases with increasing $U$ regardless of
the electron concentration\cite{Sanchez:99a}.

These are already well-known results\cite{Kawakami:89,Usuki:90}
that the approximation considered here is able to reproduce, thus
capturing the physics of the charge excitations near the
metal-insulator transition. The temperature at which the gap
closes due to the thermal excitations is somewhat larger in
\emph{COM}~1 than in the Bethe Ansatz
results\cite{Kawakami:89,Usuki:90}. This is coherent with the
larger value that we obtain for the gap. Also, $\chi_{c}$ is more
enhanced near half filling (for instance, at $U=8t$ and $n=0.8$,
$\chi_{c,max}^{\emph{COM} 1}\sim0.9$ while
$\chi_{c,max}^{BA}\sim0.25$) because of the faster rate at which
the gap opens in our approximation with respect to \emph{BA}.

Information on the physics of the spin excitations can be
extracted from the evolution of the magnetic susceptibility with
temperature and Coulomb interaction. We discuss the situation
close to half filling. The magnetic susceptibility $\chi_{s}$ is
calculated by means of Eq.~(\ref{spinsuseq}) and its behavior, as
a function of the temperature and $U$, is shown in
Fig.~\ref{Fig18}. $\chi_{s}$ presents a peak at low temperatures
which moves to lower ones as $U$ increases. It is also worth
noticing that $\chi_{s}$ is generally enhanced by $U$. At half
filling [not shown], our results indicate a renormalization of the
spin excitations, as it is obtained in the case of the charge
susceptibility. This behaviour of $\chi_{s}$ for the interacting
half-filled chain does not reproduce the \emph{BA}
result\cite{Kawakami:89,Usuki:90}. According to the exact
solution, $\chi_{s}$ should show the same qualitative behaviour
regardless of the particle density, because the metal-insulator
transition at $n=1$ does not renormalize the spin excitations.
Such disagreement between \emph{COM}~1 and \emph{BA} is due to the
computation of the two-particle Green's functions within the
one-loop approximation. In this approximation, the charge and spin
correlation functions have the same spatial dependence (cfr.
Eq.~(\ref{susCH-SP})), and consequently, gaps in charge or spin
sectors will open simultaneously. Therefore, we obtain the same
qualitative behaviour for both the charge and spin susceptibility.

This disagreement is also present in the results we have obtained
for the spin-spin correlation function (see Fig.~\ref{Fig19}). It
presents a typical paramagnetic behavior (i.e., it is always
negative at any finite distance) in contrast with the exact
diagonalization results\cite{Shiba:72a,Xu:92} which report
short-range antiferromagnetic correlations (i.e., a spin-spin
correlation function alternating in sign between sites and rapidly
decreasing). Nevertheless, the present approach preserves the
$2k_{{\rm F} }$ oscillations for any value of the Coulomb
interaction.

\subsection{Properties at arbitrary filling}

\emph{BA} predicts a non-magnetic metallic ground state for the
\emph{1D} Hubbard model at arbitrary filling, with gapless charge
and spin excitations \cite{Lieb:68}. Ground-state properties, like
the internal energy, chemical potential and local moment, were
studied following Lieb and Wu as a function of the electron
density and Coulomb interaction\cite{Shiba:72,Carmelo:88}, mainly
in the large-$U$ limit, where analytic expressions could be
obtained. By means of the formulation developed by Takahashi for
finite temperatures \cite{Takahashi:72}, some thermal properties
at arbitrary filling were also
calculated\cite{Kawakami:89,Usuki:90}. More recently, important
single-particle properties, as the spectral function and momentum
distribution function, have been evaluated within the
\emph{BA}\cite{Ogata:90,Qin:96} and by means of numerical
techniques, like \emph{qMC} \cite{Xu:92,Sorella:90,Preuss:94}.
Information on the charge and spin dynamics is the object of the
most recent studies on one-dimensional models, with the aim of
applying them to real systems. Thus, some authors have analyzed
the corresponding two-body correlation functions of the \emph{1D}
Hubbard model by means of numerical
approaches\cite{Xu:92,Sorella:90}, by using the exact \emph{BA}
solution\cite{Ogata:90,Schulz:90} or through other analytic
approaches, like \emph{g-ology} \cite{Voit:95}.

In the following sections we present the results obtained for such
quantities within the \emph{COM} for the Hubbard chain away from
half filling. The \emph{COM}~2 solution is not considered, since,
as we have commented above, it does not provide a good description
of the system in the case of half filling. Thus, the \emph{COM}~1
results are compared to the ones available by other systematics.
We devote special attention to the quarter-filled case, since it
seems to be a relevant particle density for many \emph{1D} organic
metals\cite{Vescoli:98}.

\subsubsection{Local properties}

\begin{figure}[tb]
\begin{center}
\includegraphics[width=8cm]{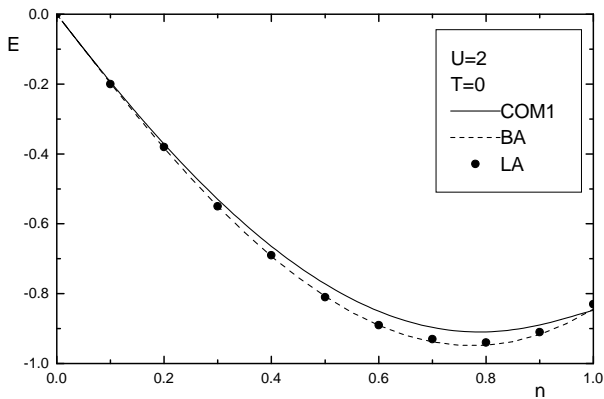}
\end{center}
\caption{Internal energy $E$ as a function of $n$ for $T=0$ and
$U=2$.} \label{Fig20}
\end{figure}

\begin{figure}[tb]
\begin{center}
\includegraphics[width=8cm]{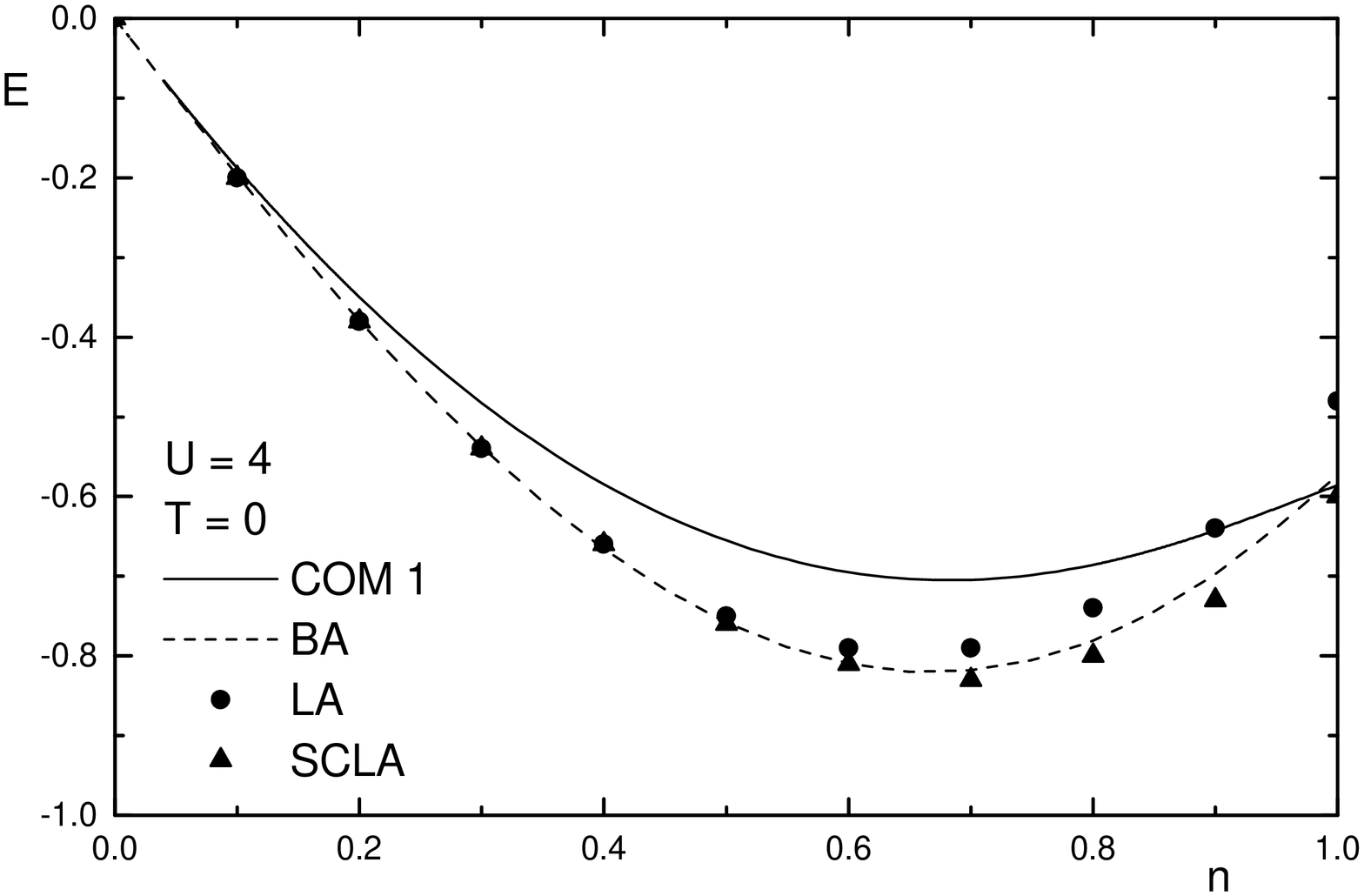}
\end{center}
\caption{Internal energy $E$ as a function of $n$ for $T=0$ and
$U=4$.} \label{Fig21}
\end{figure}

\begin{figure}[tb]
\begin{center}
\includegraphics[width=8cm]{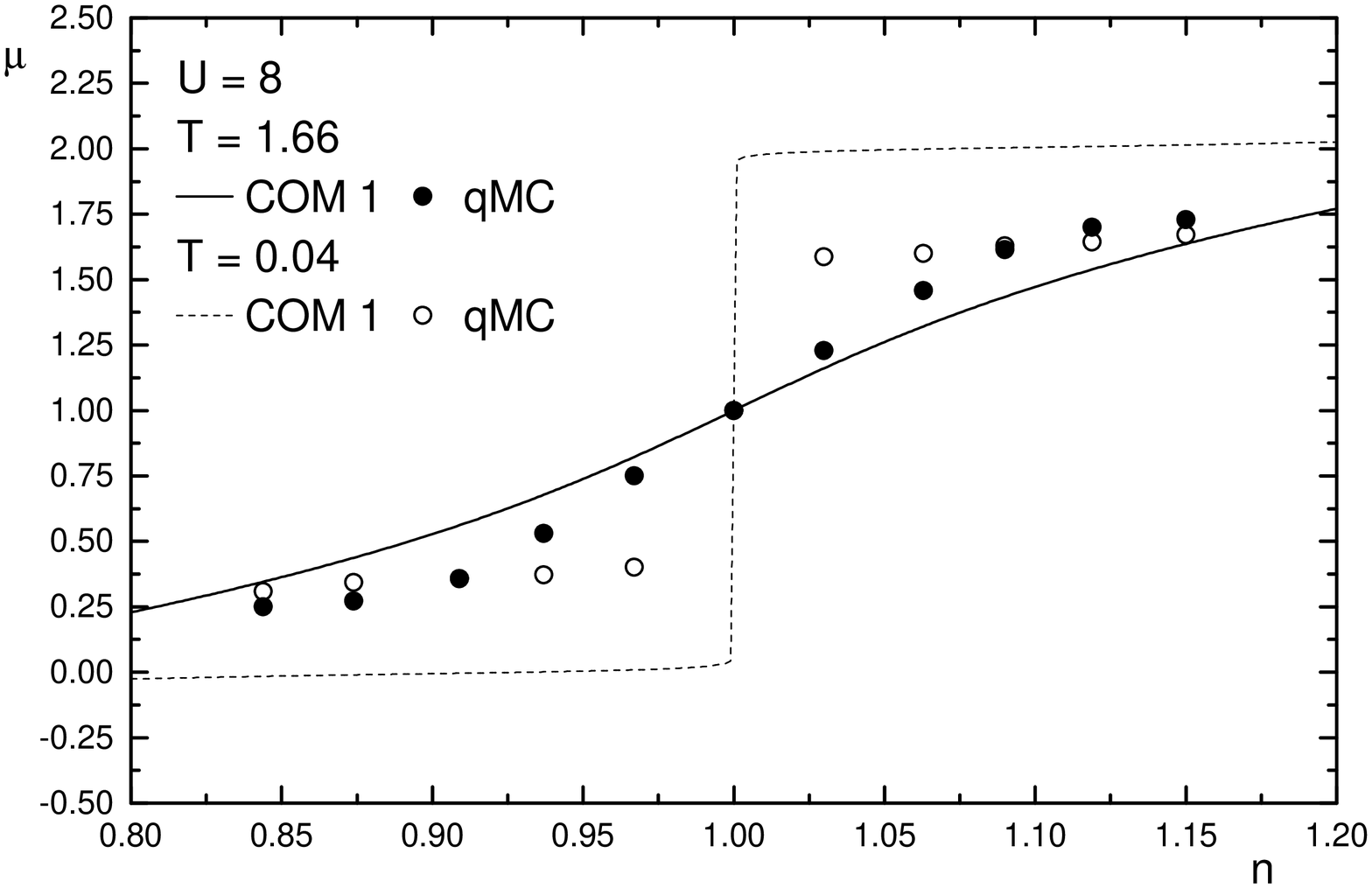}
\end{center}
\caption{Chemical potential $\mu$ as a function of $n$ for $T=0$
and $U=2$ and $4$.} \label{Fig22}
\end{figure}

\begin{figure}[tb]
\begin{center}
\includegraphics[width=8cm]{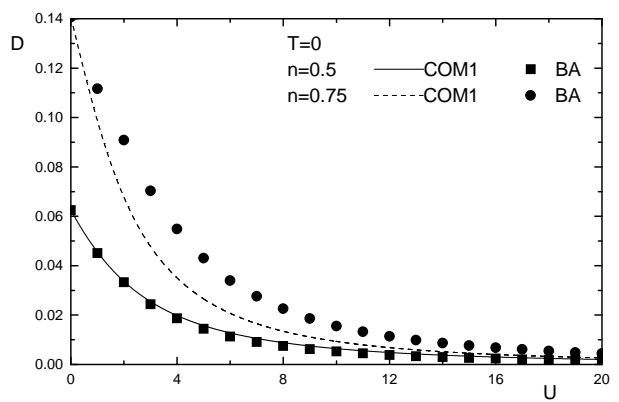}
\end{center}
\caption{Double occupancy $D$ as a function of $U$ for $T=0$ and
$n=0.5$ and $0.75$.} \label{Fig23}
\end{figure}

The doping dependence of the internal energy, calculated by means
of Eq.~(\ref{EH}) in the \emph{COM}~1 solution (i.e.,
$E_{H}^{\emph{COM}1}$), is shown in Figs.~\ref{Fig20} and
\ref{Fig21} for two values of the Coulomb interaction. For
comparison, we also report the \emph{BA} results\cite{Shiba:72}
and the ladder (\emph{LA}) and self-consistent ladder
(\emph{SCLA}) approaches\cite{Hirsch:80,Buzatu:95}. \emph{COM}~1
agrees reasonably with \emph{BA}, reaching the best agreement at
half filling. The ladder approximation\cite{Buzatu:95} deviates
more and more from the \emph{BA} as approaching half filling; the
self-consistent ladder approximation \cite{Buzatu:95} probes
excellently at any doping.

The evolution of the chemical potential with the particle density
for \emph{COM}~1 is compared in Fig.~\ref{Fig22} with some
numerical data\cite{Schulte:96}. There is a good agreement over
all the range of doping for the higher temperature. The
disagreement at the lower temperature can be understood by looking
at the size of the gap in \emph{COM}~1: \emph{COM}~1 gap is larger
than the \emph{BA} one and forces the chemical potential to assume
lower (higher) values than in the \emph{BA} solution for $n<1$
($n>1$).

The good agreement with \emph{BA}, at quarter-filling and for any
value of $U$, occurs for the double occupancy too. For other
values of filling the agreement is less satisfactory (see
Fig.~\ref{Fig23}).

\subsubsection{Thermodynamic properties}

\begin{figure}[tb]
\begin{center}
\includegraphics[width=8cm]{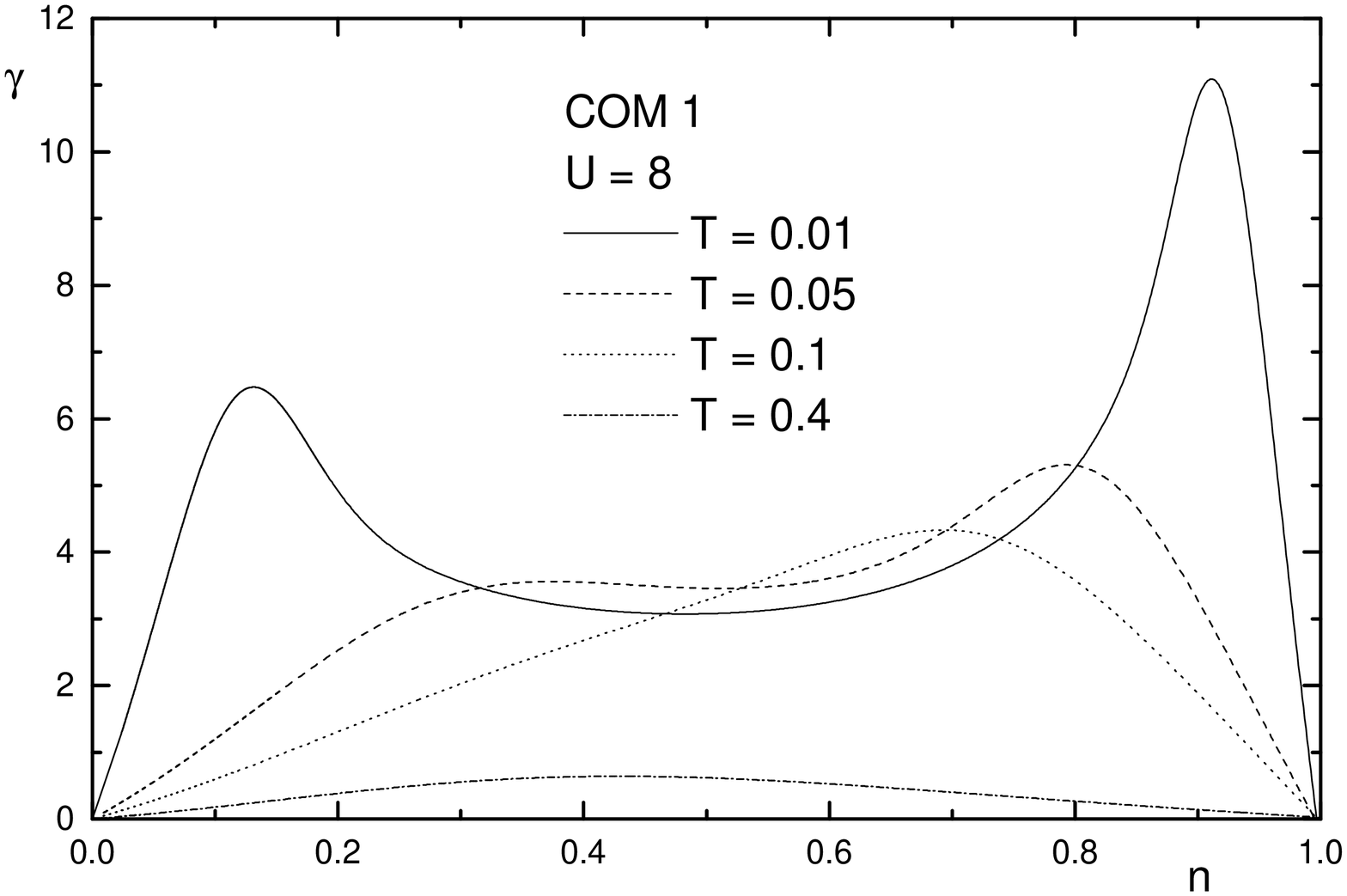}
\end{center}
\caption{$\gamma=C_{H}/T$ as a function of $n$ for $U=8$ and
$T=0.01$, $0.05$, $0.1$ and $0.4$ (\emph{COM}~1 solution).}
\label{Fig24}
\end{figure}

\begin{figure}[tb]
\begin{center}
\includegraphics[width=8cm]{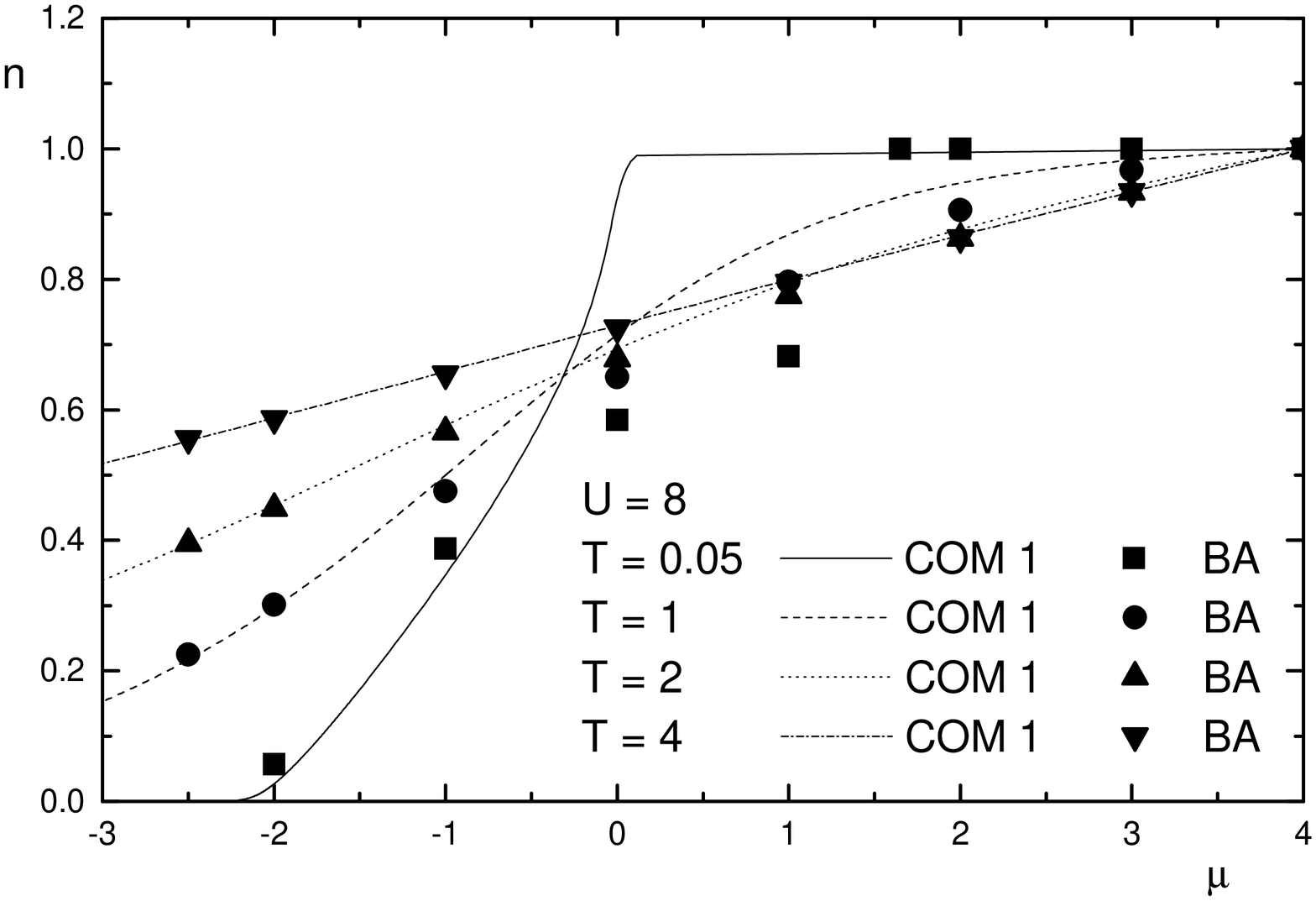}
\end{center}
\caption{Filling $n$ as a function of the chemical potential $\mu$
for $U=8$ and $T=0.05$, $1$, $2$ and $4$ (\emph{COM}~1 solution).}
\label{Fig25}
\end{figure}

The evolution of the specific heat $C_{H}$ with the particle
density is studied for various temperatures in the strong coupling
regime. As shown in Fig.~\ref{Fig24}, $\gamma=C_{H}/T$ has a peak
at low densities that simply reflects the shape of the density of
states (see discussion on single-particle properties in the next
section). As $n$ approaches half filling, another peak develops at
low temperatures. This indicates an increasing number of excited
states due to the renormalization of the charge fluctuations at
the opening of the Mott gap. Let us note that such a peak does not
appear in strongly correlated systems which do not have a
metal-insulator transition like the \emph{1D} electron gas with
delta-function interactions\cite{Usuki:89}. As one should expect,
with increasing temperature, the two peaks merge. \emph{COM}~1
results for $C_{H}/T$ recover qualitatively those obtained by
\emph{BA} \cite{Kawakami:89}. There are some quantitative
differences, namely: the first peak appears at higher fillings and
the double-peak feature survives up to higher temperatures in
\emph{BA}; the peak near half filling is higher in \emph{COM}~1.
This latter difference is expected as the \emph{COM}~1 gap is
larger than the \emph{BA} one.

To complete the above discussion we show in Fig.~\ref{Fig25} the
particle density $n$ versus the chemical potential $\mu$ for
various temperatures. \emph{COM}~1 results are compared with the
\emph{BA} ones of Ref.~\onlinecite {Kawakami:89}. The agreement is
very good at low temperatures for densities smaller than $0.55$.
In the half-filled chain, $T\sim t$ is a relevant temperature as
it signs the border between $T$-regions dominated by either spin
or charge correlations \cite{Shiba:72a,Schulte:96}. The agreement
between \emph{COM}~1 and \emph{BA} at $T\geq t$ is very good for
the whole range of filling. Of course, at higher temperatures
\emph{COM}~1 result reaches an excellent agreement with \emph{BA}
since the effect of correlations is completely suppressed.

\subsubsection{Single-particle properties}

\begin{figure}[tb]
\begin{center}
\includegraphics[width=8cm]{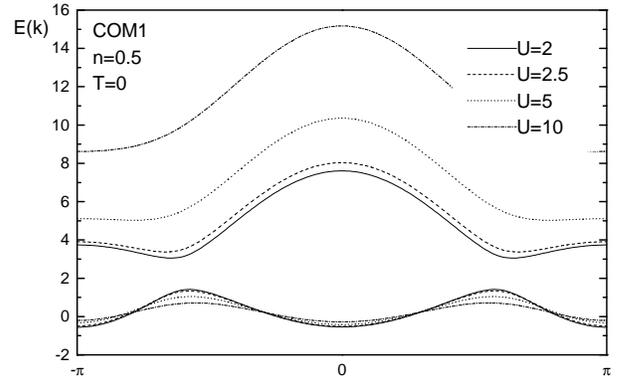}
\end{center}
\caption{Energy spectra $E(k)$ for $n=0.5$, $T=0$ and $U=2$,
$2.5$, $5$ and $10$ (\emph{COM}~1 solution).} \label{Fig27}
\end{figure}

\begin{figure}[tb]
\begin{center}
\includegraphics[width=8cm]{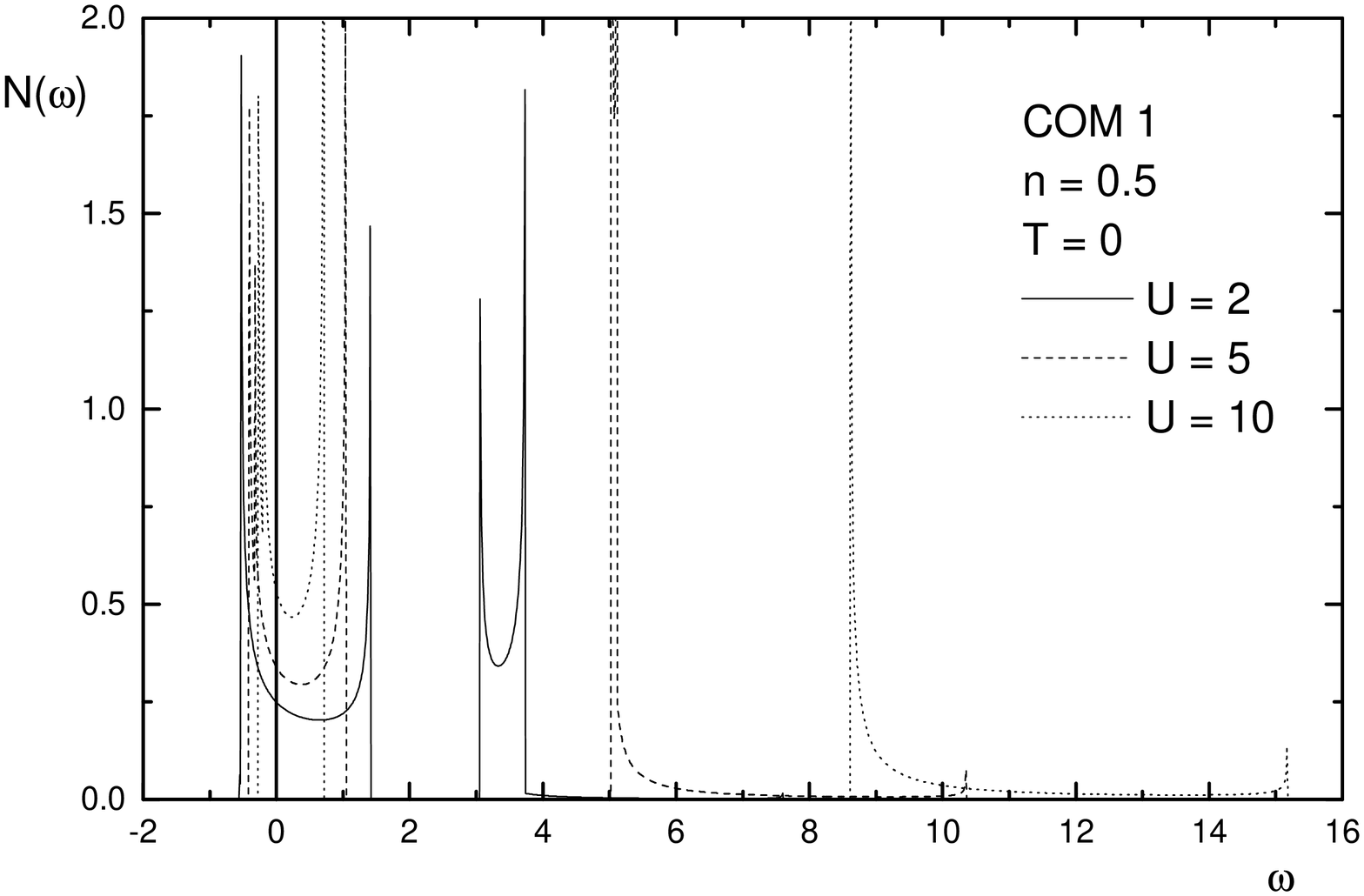}
\end{center}
\caption{Density of states for $n=0.5$, $T=0$ and $U=2$, $5$ and
$10$ (\emph{COM}~1 solution).} \label{Fig28}
\end{figure}

\begin{figure}[tb]
\begin{center}
\includegraphics[width=8cm]{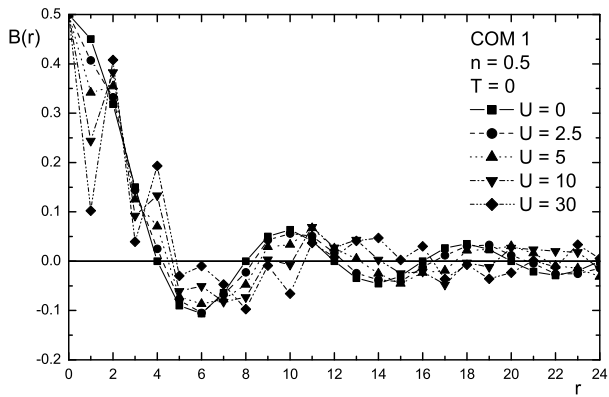}
\end{center}
\caption{Momentum distribution function in real space $B(r)$ for
$n=0.5$, $T=0$ and $U=0$, $2.5$, $5$, $10$ and $30$ (\emph{COM}~1
solution).} \label{Fig29}
\end{figure}

The evolution of the band structure with the interaction strength
$U$ is remarkable. Away from half filling the \emph{AF}-band
pattern of the upper Hubbard subband in \emph{COM}~1 disappears as
$U$ increases; the first electron excitations appear at $k=\pm\pi$
and the first hole excitations move slightly away from the
half-filling position $k=\pm\pi/2$ (see Fig.~\ref{Fig27}).

The corresponding density of states is shown in Fig.~\ref{Fig28}.
It is well known that the density of states in the non-interacting
case exhibits two van Hove singularities at the edges of the
non-interacting band. In the interacting case each of these
singularities splits in two since we have two distinct Hubbard
subbands. Due to the \emph{AF}-like band shape, three van Hove
singularities appear in each subband leading to the six peaks of
the figure. The third structure in the upper subband of
Fig.~\ref{Fig28} is smoothed because, away from half filling, as
$U$ increases the \emph{AF}-band shape of the excitation spectrum
disappears.

As we can deduce from the above results the system, away from half
filling, is a conductor for any value of the interaction strength
$U$ in agreement with the \emph{BA} results. Then, a natural
question arises: what universality class does this
\emph{conductor} belong to? This is a central issue in the physics
of \emph{1D}-models.

The momentum distribution function $n(k)$ is a relevant property
because from its behavior the Fermi-liquid or non-Fermi liquid
nature of the excitations can be inferred. As it is well known, a
finite jump in the zero-temperature momentum distribution function
at the Fermi momentum $k_{{\rm F}}$ would indicate that the
quasiparticle excitations can be described by a Fermi liquid. Some
variational\cite{Fazekas:88,Yokohama:87},
analytic\cite{Carmelo:88} and numerical\cite{Hirsch:84} approaches
failed in understanding the nature of the discontinuity. More
recently, approaches valid in the weak coupling regime
(\emph{g-ology} and \emph{qMC}) have found a power-law singularity
at $k_{{\rm F}}$ predicting a marginal Fermi-liquid nature away
from half filling\cite{Sorella:90}. The power-law coefficient
seems to be an increasing function of $U$ and a decreasing
function of the particle density $n$.

The approximation that we use limits the information that can be
obtained from the momentum distribution function. Namely, any
two-pole approximation has a momentum distribution function with
sharp discontinuities at the values of momentum where the Hubbard
subbands cross the Fermi level. Despite this strong limitation,
the momentum distribution function obtained at quarter filling in
\emph{COM}~1 presents, besides a sharp jump at $k_{{\rm F}},$
another discontinuity near $3k_{{\rm F}}$ . This latter feature is
also obtained, as a weak singularity, in \emph{BA} calculations in
the large-$U$ limit\cite{Ogata:90}.

The momentum distribution function in real space obtained in
\emph{COM}~1 is shown in Fig.~\ref{Fig29} for various interaction
strengths and quarter filling. In the weak coupling regime, the
distribution function has nodes at $r=4a,8a,12a,\ldots$ which
correspond to an oscillation of wavelength $8a$. Since for these
values of interaction and particle density the Fermi momentum is
$\frac{\pi}{4}$, this is just a $k_{{\rm F}}$ oscillation. Such an
oscillation is also observed by exact diagonalization
results\cite{Xu:92} for small $U$. For stronger interactions, a
weak incommensurate modulation, in addition to the $k_{{\rm F}}$
oscillation, appears and corresponds to a $3k_{{\rm F}}$
oscillation again in agreement with the exact diagonalization
calculations\cite{Xu:92}. The origin of this $3k_{{\rm F}}$
feature is however not clear. According to other numerical
calculations\cite{Qin:96} the $3k_{{\rm F}}$ singularity seems to
be a finite-size effect and vanishes in the thermodynamic limit.
We actually confirm its presence also in the bulk system.

\subsubsection{Correlation functions and susceptibilities}

\begin{figure}[tb]
\begin{center}
\includegraphics[width=8cm]{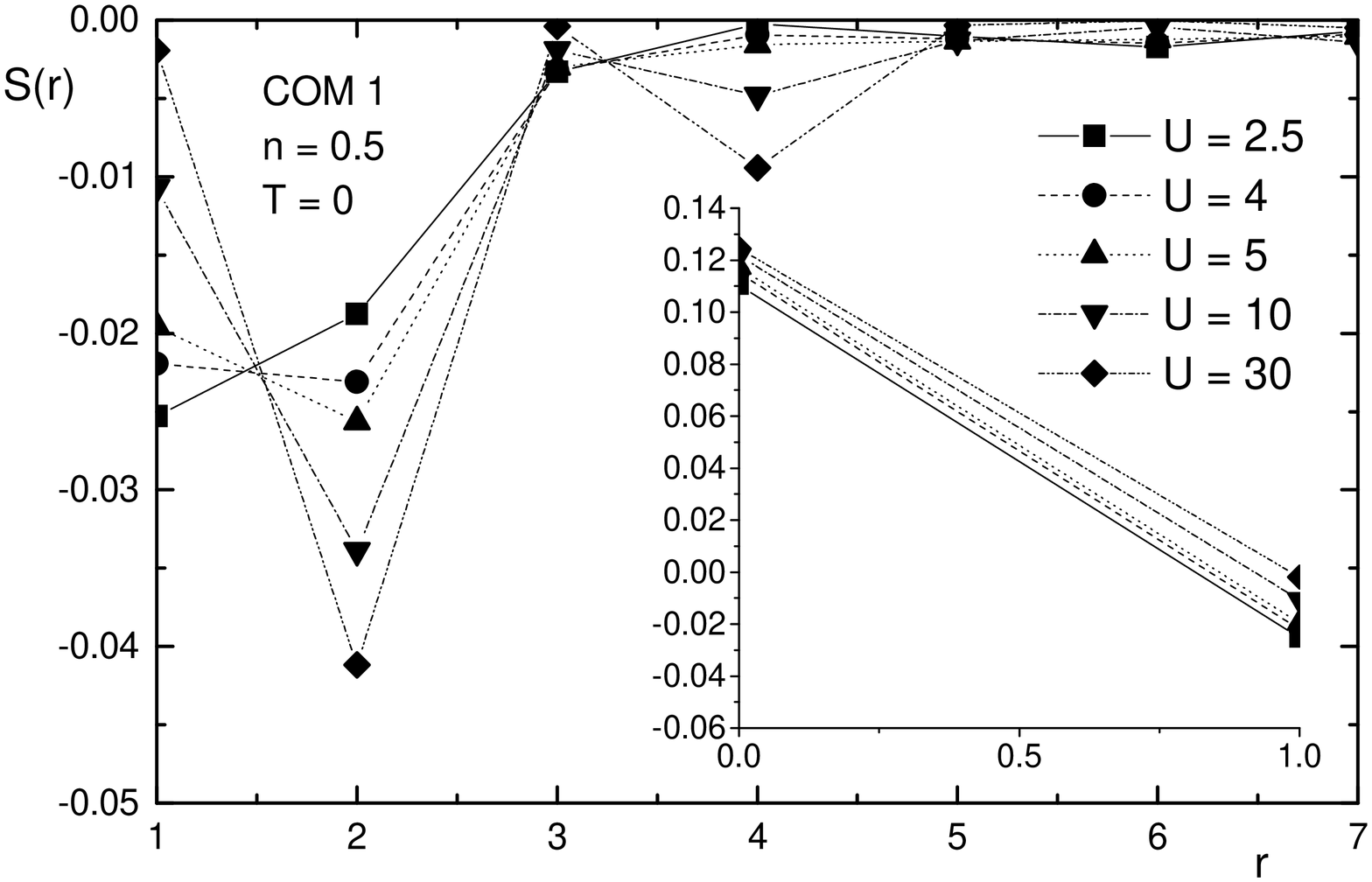}
\end{center}
\caption{Spin correlation function $S(r)$ for $n=0.5$, $T=0$ and
$U=2.5$, $4$, $5$, $10$ and $30$ (\emph{COM}~1 solution).}
\label{Fig30}
\end{figure}

\begin{figure}[tb]
\begin{center}
\includegraphics[width=8cm]{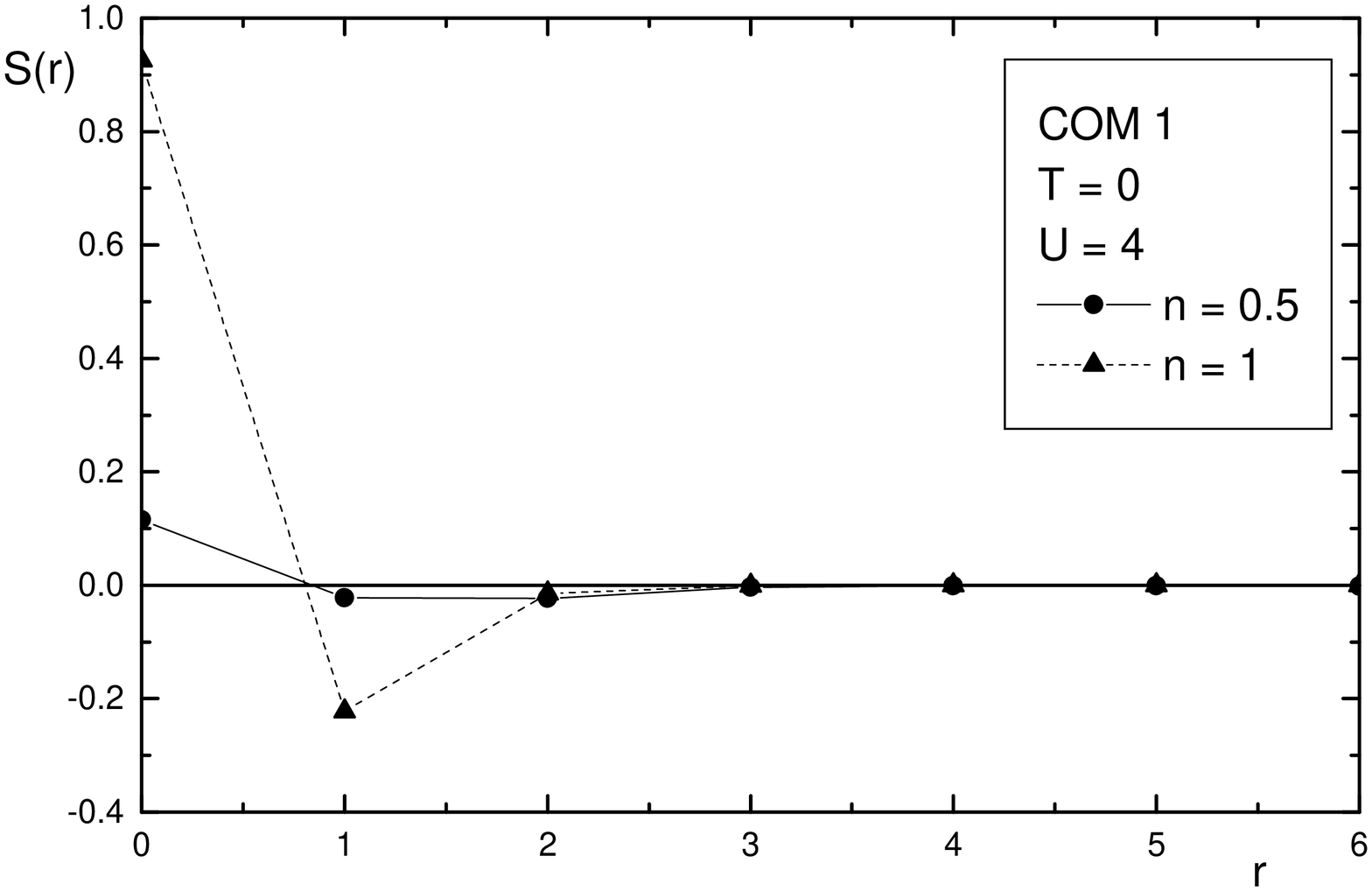}
\end{center}
\caption{Spin correlation function $S(r)$ for $U=4$, $T=0$ and
$n=0.5$ and $1$ (\emph{COM}~1 solution).} \label{Fig31}
\end{figure}

\begin{figure}[tb]
\begin{center}
\includegraphics[width=8cm]{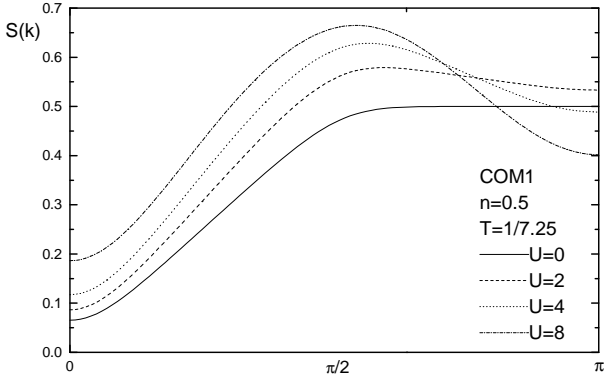}
\end{center}
\caption{Spin correlation function $S(k)$ for $n=0.5$, $T=0$ and
$U=0$, $2$, $4$ and $8$ (\emph{COM}~1 solution).} \label{Fig32}
\end{figure}

\begin{figure}[tb]
\begin{center}
\includegraphics[width=8cm]{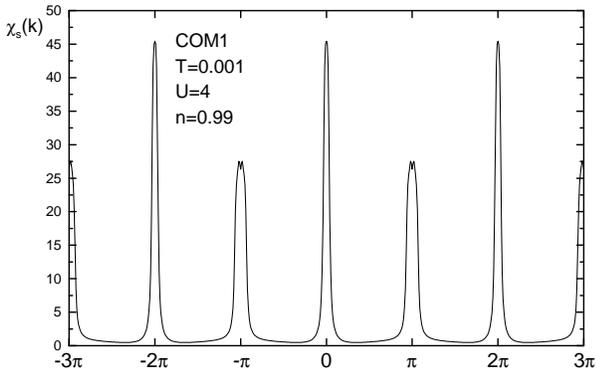}
\end{center}
\caption{Spin susceptibility $\chi_{s}(k)$ for $U=4$, $T=0.001$
and $n=0.99$ on the extended zone (\emph{COM}~1 solution).}
\label{Fig33}
\end{figure}

\begin{figure}[tb]
\begin{center}
\includegraphics[width=8cm]{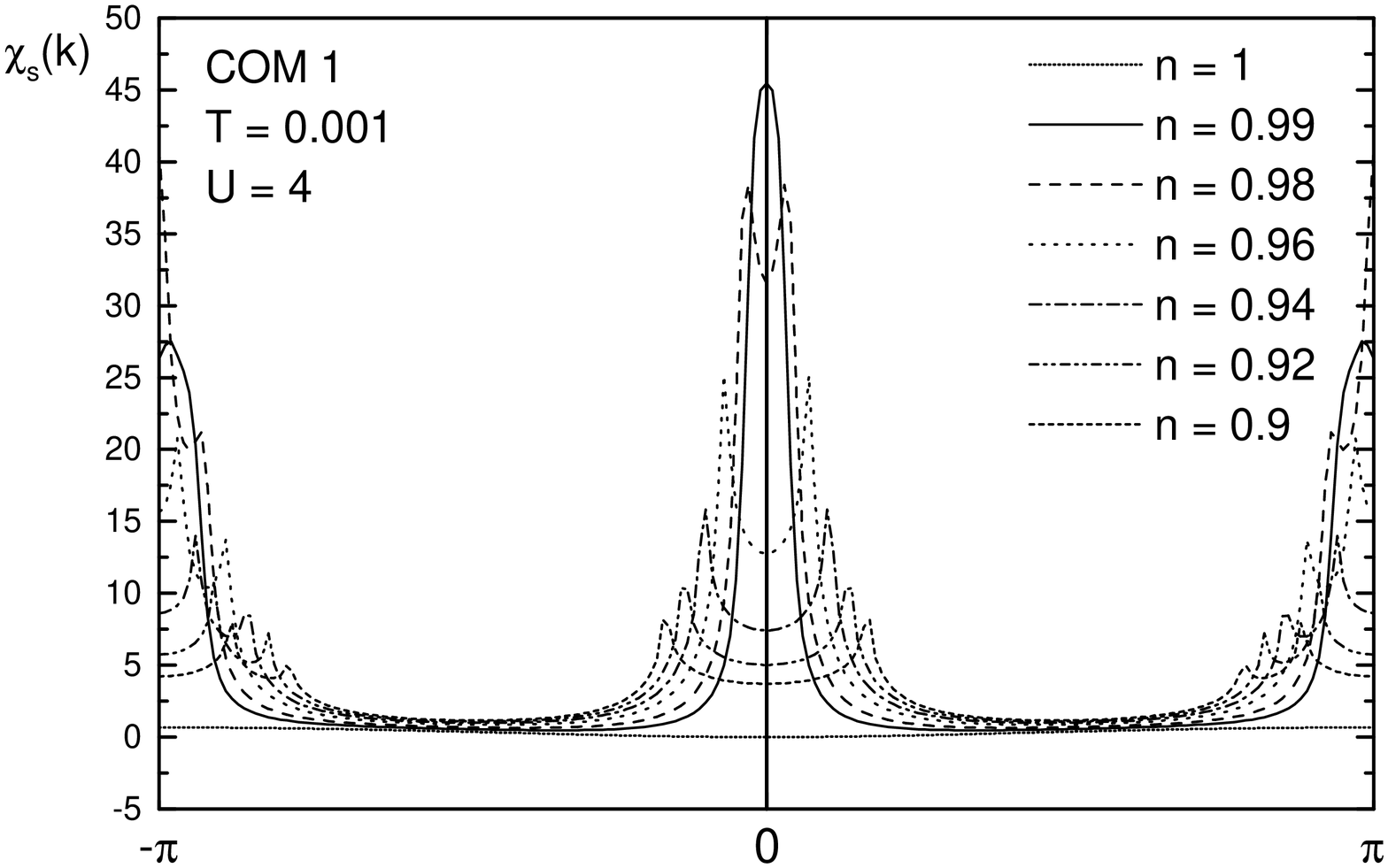}
\end{center}
\caption{Spin susceptibility $\chi_{s}(k)$ for $U=4$, $T=0.001$
and $n=1$, $0.99$, $0.98$, $0.96$, $0.94$, $0.92$ and $0.9$
(\emph{COM}~1 solution).} \label{Fig34}
\end{figure}

\begin{figure}[tb]
\begin{center}
\includegraphics[width=8cm]{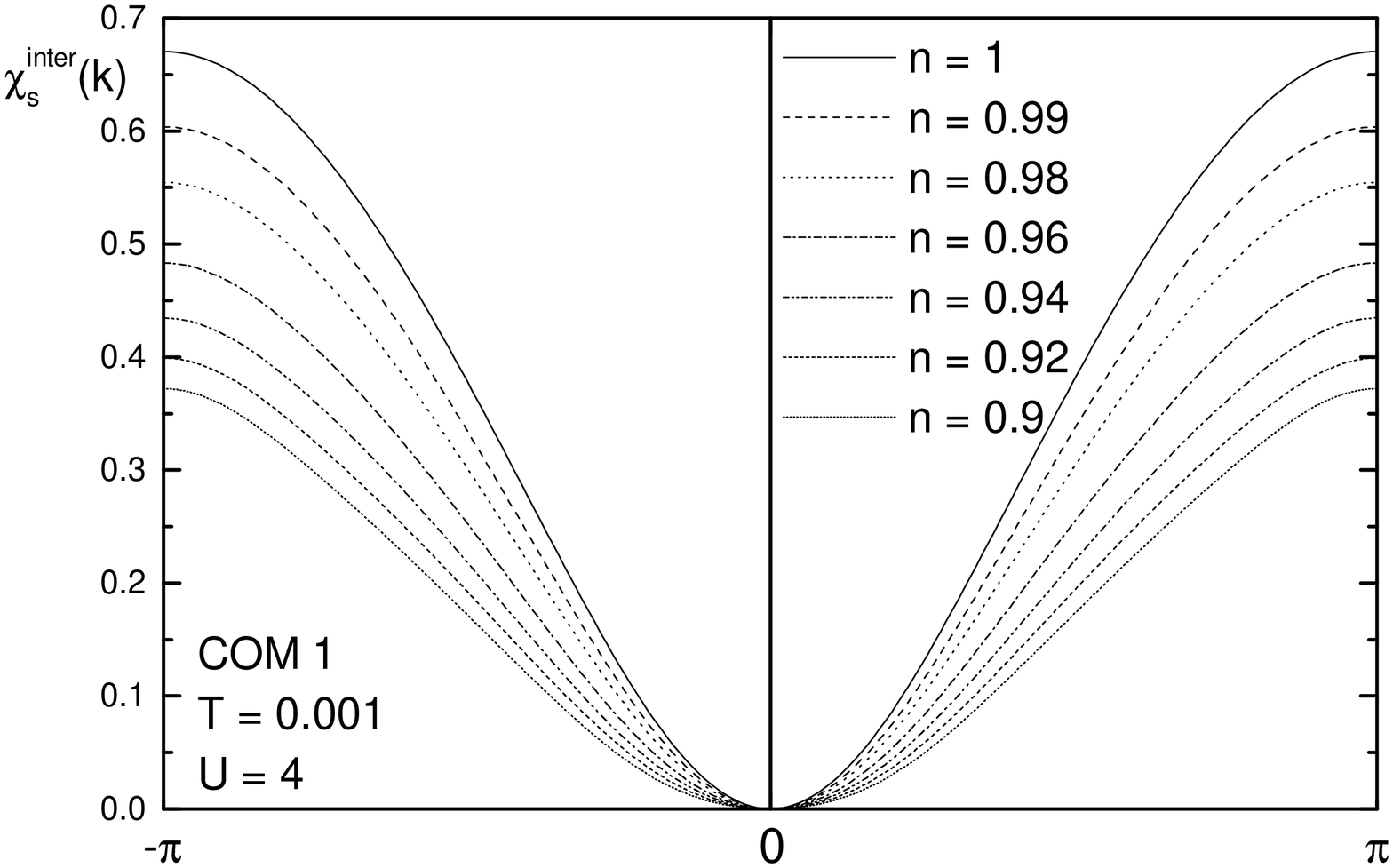}
\end{center}
\caption{Interband spin susceptibility $\chi^{inter}_{s}(k)$ for
$U=4$, $T=0.001$ and $n=1$, $0.99$, $0.98$, $0.96$, $0.94$, $0.92$
and $0.9$ (\emph{COM}~1 solution).} \label{Fig35}
\end{figure}

\begin{figure}[tb]
\begin{center}
\includegraphics[width=8cm]{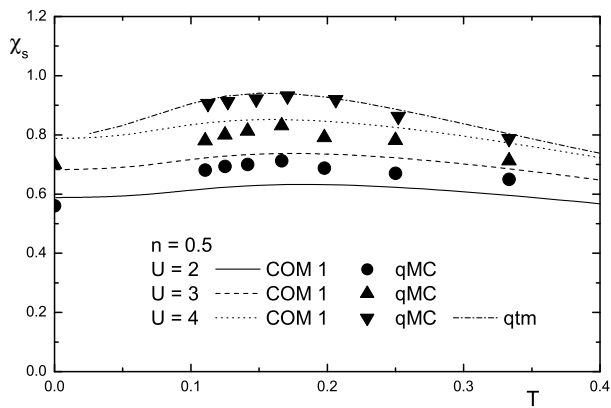}
\end{center}
\caption{Static and uniform spin susceptibility $\chi_{s}$ as
function of $T$ for $n=0.5$ and $U=2$, $3$ and $4$.} \label{Fig36}
\end{figure}

\begin{figure}[tb]
\begin{center}
\includegraphics[width=8cm]{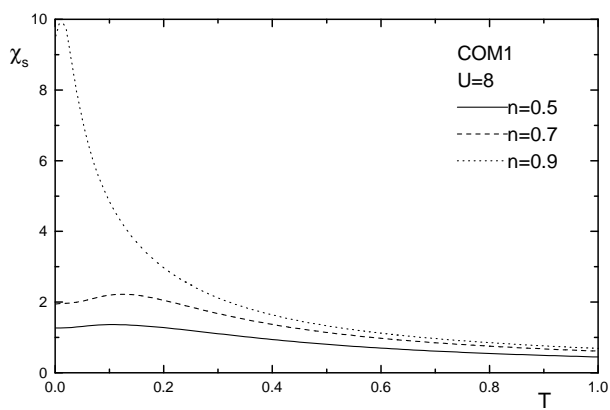}
\end{center}
\caption{Static and uniform spin susceptibility $\chi_{s}$ as
function of $T$ for $U=8$ and $n=0.5$, $0.7$ and $0.9$
(\emph{COM}~1 solution).} \label{Fig37}
\end{figure}

\begin{figure}[tb]
\begin{center}
\includegraphics[width=8cm]{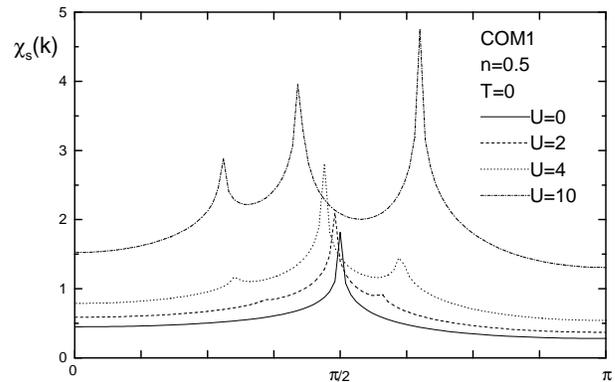}
\end{center}
\caption{Spin susceptibility $\chi_{s}(k)$ for $n=0.5$, $T=0$ and
$U=0$, $2$, $4$ and $10$ (\emph{COM}~1 solution).} \label{Fig38}
\end{figure}

\begin{figure}[tb]
\begin{center}
\includegraphics[width=8cm]{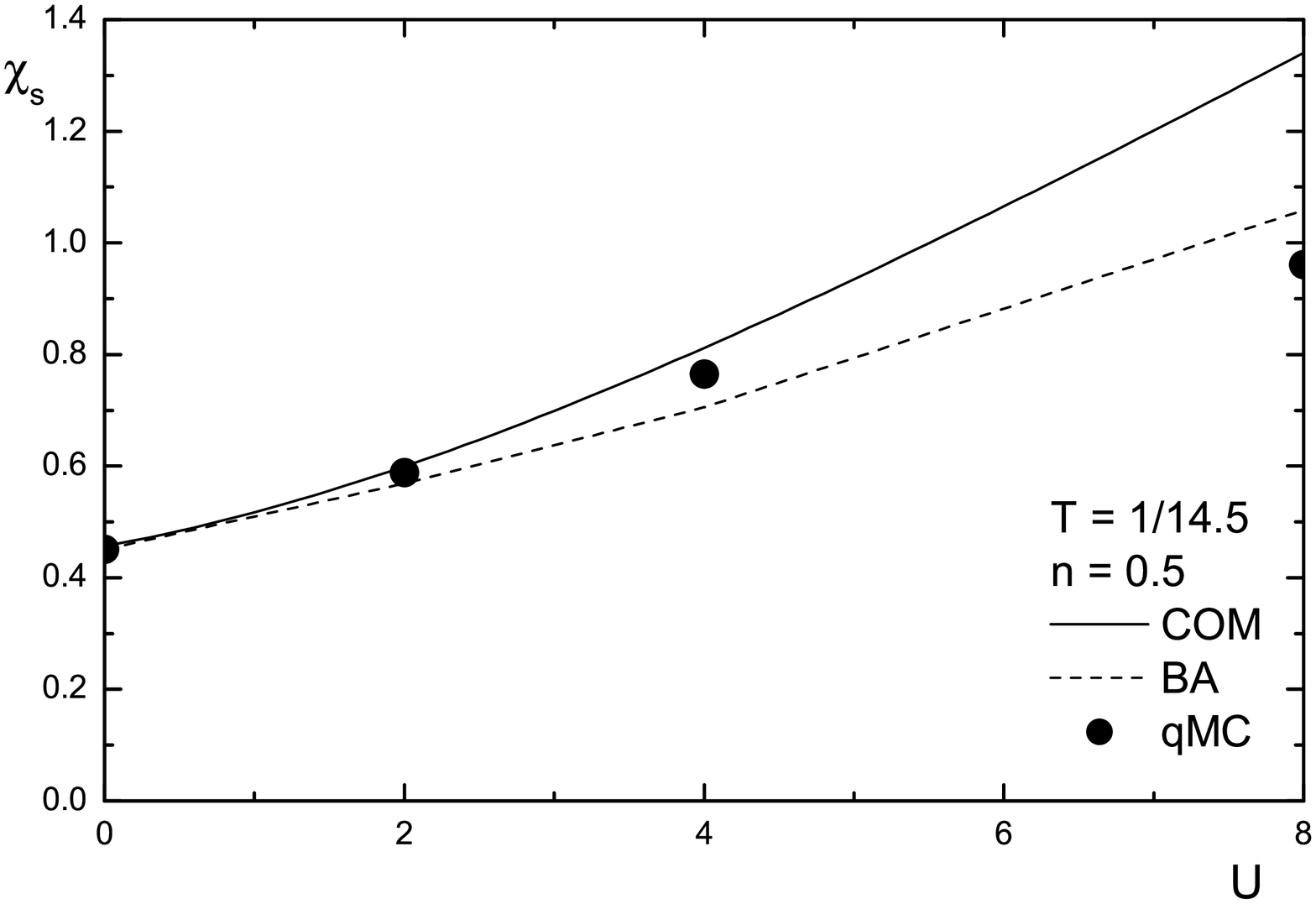}
\end{center}
\caption{Static and uniform spin susceptibility $\chi_{s}$ as a
function of $U$ for $n=0.5$ and $T=1/14.5$.} \label{Fig39}
\end{figure}

\begin{figure}[tb]
\begin{center}
\includegraphics[width=8cm]{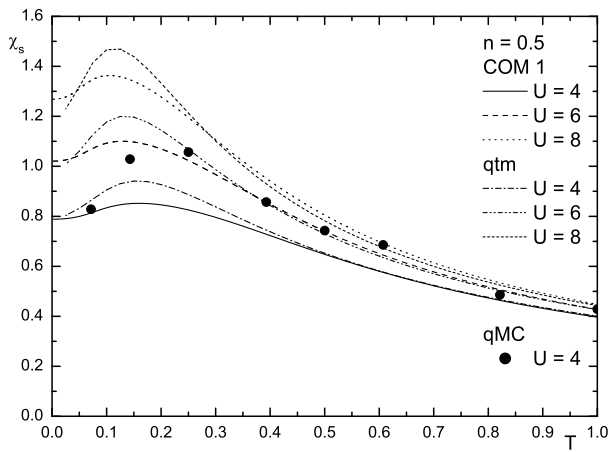}
\end{center}
\caption{Static and uniform spin susceptibility $\chi_{s}$ as a
function of $T$ for $n=0.5$ and $U=4$.} \label{Fig40}
\end{figure}

\begin{figure}[tb]
\begin{center}
\includegraphics[width=8cm]{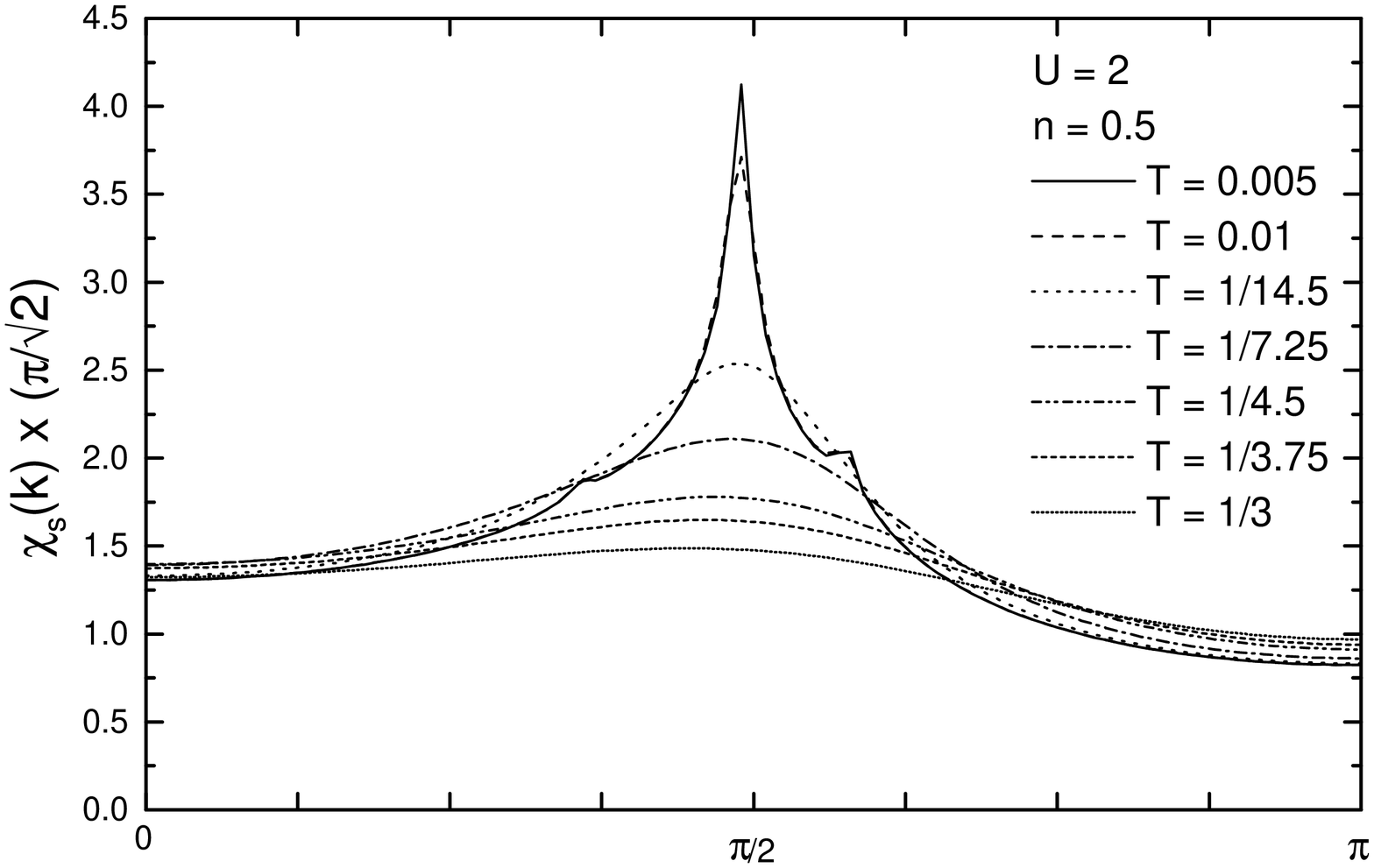}
\end{center}
\caption{Normalized spin susceptibility
$\chi_{s}(k)\frac\pi{\sqrt{2}}$ for $n=0.5$, $U=2$ and $T=0.005$,
$1/14.5$, $1/7.25$, $1/4.5$, $1/3.75$ and $1/3$.} \label{Fig41}
\end{figure}

The spin correlation function gives information about the spin
dynamics of the system. Combining this information with the one
obtained from the momentum distribution function, we can analyze
how the electron dynamics is affected by the surrounding spins,
since the spin configuration is modified when an electron moves to
another site. Therefore, the electron motion will depend on how
the spin configuration constrains the electron dynamics.

The spin correlation function in real space is evaluated in
\emph{COM 1} for various interacting regimes and relevant particle
densities. The results are shown in Figs.~\ref{Fig30} and
\ref{Fig31}. As we can see, the amplitude of spin correlation
increases with $U$ and is smeared out away from half filling. The
larger the Coulomb interaction is, the faster the correlations
decay. The physical picture that emerges for the half-filled and
quarter-filled Hubbard chain is different. At $n=1$ there are
strong antiferromagnetic spin correlations that decay very fast
(at second neighbor sites they are almost zero). The existence of
short-range AF order is also visible in the band spectrum as we
previously discussed in detail. For $n=0.5$ these correlations are
much weaker and decay more slowly than in the half-filled case.
These results agree with the ones described in
Ref.~\onlinecite{Xu:92} using exact diagonalization techniques. In
these calculations\cite{Xu:92} the spin correlation function has a
$2k_{{\rm F}}$ oscillation that is not smeared out away from half
filling. Such oscillation is not observed in Figs.~\ref{Fig30} and
\ref{Fig31} due to the very fast decay of the correlation
amplitude, but it appears as a $2k_{{\rm F}}$ feature in the
momentum-dependent spin correlation function $S(k)$ (see
Fig.~\ref{Fig32} and explanation below).

The spin correlation function in $k$ space has been recently
studied by means of various approaches. For instance, by
\emph{qMC} in the weak interacting
regime\cite{Hirsch:83,Imada:89}, by means of \emph{BA} in the
strong interacting regime\cite{Ogata:90} and for any value of $U$
by exact diagonalization technique\cite{Xu:92}. All of them find a
very narrow $2k_{{\rm F}}$ peak which is incommensurate away from
half filling. These calculations were performed on finite-size
systems. A detailed study of the size dependence of $S(2k_{{\rm
F}})$ is given in Ref. \onlinecite{Ogata:90}, where it is shown
how the peak narrows and increases with the system size.

The spin-correlation function in $k$ space is shown in
Fig.~\ref{Fig32} for the \emph{COM}~1 solution at quarter filling
and various interaction strengths. As $U$ increases, an evolution
towards a peaked curve at $2k_{{\rm F}}$ is observed, in
qualitative agreement with previous
calculations\cite{Xu:92,Ogata:90,Hirsch:83,Imada:89}. We must,
however, remark that within the present approach the \emph{peak}
at $2k_{{\rm F}}$ is much broader. At half filling, the height of
the $2k_{{\rm F}}$ peak is very much enhanced, in agreement with
\emph{BA} calculations in the large-$U$ limit. The reduction of
the $S(2k_{{\rm F}})$ peak as $n$ decreases from half filling is
due to the presence of holes moving in the system.

The magnetic or spin susceptibility $\chi_{s}$ gives information
about the physics of the spin excitations. This property is
calculated by means of Eq.~(\ref{spinsuseq}) and analyzed as a
function of temperature, momentum and interaction.

Close to half filling we have a paramagnetic solution with period
$2\pi$, but a strong \emph{AF} order, with a quasi-halved
Brillouin zone, is present (see Fig.~\ref{Fig33}). When we move
away from half filling the central peak opens in two separate
peaks (see Fig.~\ref{Fig34})). The incommesurability amplitude
increases linearly with doping with a coefficient of $\pi/2$ at
zero temperature. If we distinguish the intraband and interband
contributions we can show that the interband contribution is very
little (see Fig.~\ref{Fig35}). At half filling the susceptibility
is strongly reduced and goes to zero at zero temperature: the
interband contribution goes much slower to zero with temperature
and for the analyzed temperature ($T=0.001$) is the only one
present (see Figs.~\ref{Fig34} and \ref{Fig35}).

Fig.~\ref{Fig36} shows the temperature evolution of $\chi_{s}$ for
weak and intermediate coupling. Monte Carlo data\cite{Nelisse:99}
and \emph{qtm} results\cite{Juttner:98} are included for
comparison. The agreement is good, with a maximum deviation of
15\%. We shall note that the position of the maximum does hardly
depend on the strength of interaction and is located around $0.2$
(in units of $t$). This indicates that the energy necessary to
excite the spin modes does not depend on $U$. Since $U$ is of
Coulomb origin and affects the charge degrees of freedom, the
$U$-independence of the peak in the spin susceptibility shows that
the spin excitations are independent of the charge. This is
coherent with a Luttinger liquid description of the quarter-filled
Hubbard chain. On the contrary, at half filling the position of
the $T$-dependent spin susceptibility does strongly vary with $U$
(see Fig.~\ref{Fig18} and Ref.~\onlinecite{Shiba:72a}), indicating
the breaking of the Luttinger liquid picture.

Fig.~\ref{Fig37} shows the strong coupling spin susceptibility for
several particle densities. Once again, \emph{COM}~1 results at
quarter filling are in good agreement with
\emph{BA}\cite{Kawakami:89,Usuki:90} (see value and position of
$\chi_{s}^{{\rm max}}$), whereas it severely disagrees in the case
of higher particle-densities.

The low temperature, momentum dependence of the magnetic
susceptibility $\chi_{s}(k)$ at quarter filling is shown in
Fig.~\ref{Fig38} for all coupling regimes. As $U$ increases, the
peak moves towards lower $k$, indicating that spin excitations of
larger wavelength mostly contribute. Also, two satellite
structures appear and increase with $U$. Such structures have
their origin in the van Hove singularities of the density of
states. They are not observed in \emph{qMC} studies probably
because they exist only at very small temperatures (see
Fig.~\ref{Fig41}).

The static uniform susceptibility is plotted in Figs.~\ref{Fig39}
and \ref{Fig40} as function of $U$ and $T$, respectively. In
Fig.~\ref{Fig39} we also report the quantum Monte Carlo
data\cite{Hirsch:84} and the \emph{BA} results. The agreement, in
particular in the weak coupling regime, is quite good. In
Fig.~\ref{Fig40} the comparison with the \emph{qMC}
\cite{Hirsch:84} and the \emph{qtm}\cite{Juttner:98} results
shows: 1) the \emph{qMC} data do not agree at all with the
\emph{qtm} results, which are practically exact; 2) a good
qualitative agreement between \emph{COM} and \emph{qtm} at very
low temperatures for all three values of Coulomb repulsion $U$(
the \emph{qtm} results present a more pronounced peak) and a very
good quantitative agreement at intermediate and high temperatures.
Anyway, the position of the peak is very well reproduced
confirming once more the capability of the present method to catch
the spin energy scale although the retained spin correlations are
weaker than what is expected according to the exact and numerical
results.

The weak coupling spin susceptibility (normalized to its
non-interacting value at $k=0$:
$\chi_{s}(0)[U=0]=\frac{\sqrt{2}}\pi$, see Eq.~\ref{ChisU0}) is
shown in Fig.~\ref{Fig41}. Despite the simplicity of the one-loop
approximation we get a good qualitative agreement with the quantum
Monte Carlo data\cite{Hirsch:84}, as opposed to \emph{RPA} fits,
which need to take different values for the renormalized
interaction, as it is remarked in Ref.~\onlinecite{Nelisse:99}. We
exactly reproduce the peak at $\pi/2$, which vanishes on
increasing temperature, and the asymmetry they found in the
intensity between $k=0$ and $k=\pi$.

\section{Conclusions}

\label{Conclusions}

In this paper we have analyzed the adequacy of the \emph{COM} to
describe the physics of correlated electrons in \emph{1D} systems;
in particular, we studied the \emph{1D} Hubbard model. Various
physical properties, like ground-state and single-particle
properties, the thermodynamics, susceptibilities and some
correlation functions, have been calculated and compared to
results obtained by means of the Bethe Ansatz, when available, and
other analytic and numerical techniques (\emph{qMC, GA, SCLA, RG,
g-ology}).

By considering a two-pole approximation and a paramagnetic state,
the model is solved within the \emph{COM}. Two mathematical
solutions (\emph{COM}~1 and \emph{COM}~2) are obtained. In the
case of half filling an analysis of the parameters and the energy
spectra that characterize the solution shows that only
\emph{COM}~1 is consistent with the \emph{BA} results. Hence, the
subsequent properties are discussed only for this solution. The
half-filled and arbitrary-filled cases are addressed separately
since they lead to different characteristics.

The essential physics of the half-filled Hubbard chain is
reproduced satisfactorily by the \emph{COM}~1 solution. A gap
opens for any finite value of the $U$ interaction and the bands
show the characteristic \emph{AF}-like features, with bandwidth of
the order of the antiferromagnetic exchange interaction
$J=\frac{4t^{2}}{U}$. This indicates, therefore, an insulator with
short-range \emph{AF} correlations, as it is known from the exact
\emph{BA} result. Also, the evolution of the total energy, the
double occupancy and the local magnetization with $U$ are in
excellent agreement with \emph{BA} and improves significantly the
results of other analytic approaches.

The thermodynamic properties give the following picture. As the
interaction strength increases relative to the non-interacting
bandwidth $W$ and reaches $U>W=4t$, two energy scales manifest in
the system in the form of two peaks in the specific heat. These
peaks are located at low ($T\sim J$) and high $T$ and are
associated to spin and charge excitations, respectively. The spin
nature of the low-$T$ peak in $C_{T}$ is confirmed by the
evolution of the local magnetization with temperature. In the
strong interacting regime, the spin and charge degrees of freedom
also manifest separately in the entropy, where $T\sim t$ is the
border between the two regions. This picture agrees qualitatively
and quantitatively (position and height of peaks in $C_{T}$ and
border between the spin and charge-dominated regions in $S_{T}$)
with \emph{BA}.

The behaviour of the charge excitations near the metal-insulator
transition is also very well captured. This can be seen from the
analysis of the charge susceptibility $\chi_{c}$ versus $T$ for
particle densities approaching $n=1$. The agreement with \emph{BA}
is only qualitative because the faster opening of the gap in
\emph{COM}~1 leads to larger values of $\chi_{c}$.

The physics of the spin excitations, extracted from the evolution
of the spin susceptibility $\chi_{s}$, is not properly described.
Our results for $\chi_{s}$ indicate, in disagreement with
\emph{BA}, a renormalization of the spin excitations as the system
approaches half filling, analogous to what is observed for the
charge excitations. This failure is inherent to the one-loop
approximation; within the latter gaps in charge and spin sectors
open simultaneously since the charge and spin correlation
functions have the same spatial dependence.

The basic physics of the arbitrary-filled Hubbard chain is
reproduced satisfactorily by the \emph{COM}~1 solution. In
agreement with \emph{BA} results, we obtain a non-magnetic metal
for any coupling regime. Namely, the band spectrum is gapless for
any value of $U$, it loses gradually its \emph{AF}-like
characteristics, and the amplitude of the spin correlation
function is much reduced in comparison with that obtained at half
filling.

Our analysis for arbitrary filling is centered mostly on $n=0.5$
which is relevant to real quasi-\emph{1D} systems. At this
particular filling, low-temperature local quantities like the
double occupancy and the chemical potential are in very good
agreement with \emph{BA} results for both weak and strong coupling
regimes. The internal energy versus particle density in
\emph{COM}~1 has a reasonable agreement with \emph{BA} although it
is not so good as that of other approaches.

The evolution of the specific heat with particle density
reproduces qualitatively the \emph{BA} results. The $T$-dependent
spin susceptibility at quarter filling has a reasonable agreement
in the weak and strong coupling regime with quantum Monte Carlo
and \emph{BA} data, respectively.

The momentum distribution function $n(k)$, the correlation
functions and the susceptibilities provide information on the
universality class this system belongs to and on its charge and
spin dynamics.

We can grasp some characteristic features. Namely, the weak
singularity at $3k_{{\rm F}}$ obtained in the large-$U$ limit
\emph{BA} calculations at quarter filling would correspond to the
discontinuity near $3k_{{\rm F}}$ that is observed in the
\emph{COM}~1 momentum distribution function. This feature would
also agree with the $3k_{{\rm F}}$ oscillation of the distribution
function in real space $C(r)$ obtained by numerical techniques.
The $k_{{\rm F}}$ oscillations of $C(r)$ observed by exact
diagonalization calculations are also reproduced in the weak
coupling \emph{COM}~1 results.

The approach manages to grasp the different physics for the
half-filled and arbitrary-filled case, in particular quarter
filling. By comparing the amplitude of the spin correlation
function in real space $S(r)$, we conclude that at quarter filling
these correlations are much weaker and decay more slowly than at
half filling, in agreement with exact diagonalization results.

The $2k_{{\rm F}}$ singularity of the $k$-dependent spin
correlation function $S(k)$ obtained in recent \emph{BA} and
numerical approaches is qualitatively described in \emph{COM}~1 at
quarter filling, where, by increasing $U$, an evolution towards a
peak structure in $S(k)$ is observed near $2k_{{\rm F}}$. This
peak is much enhanced at half filling, in agreement with
\emph{BA}.

To summarize, when integral properties are addressed (local
quantities, thermodynamics, total energy), the \emph{COM} in the
two pole approximation is accurate enough to yield a correct
description of the system. The agreement with \emph{BA} is indeed
excellent in the case of half filling and improves the results of
other analytical methods which are more lengthy and more complex
in many cases. The charge susceptibility is also well described as
the charge excitations are dominated by the energy scale set by
the opening of the Mott gap, and this is indeed caught by a
two-pole approach. However, regarding the spin dynamics the
one-loop approximation can only grasp some general physics but it
is too simple to investigate it properly. We expect to receive
better results whenever we will set up an approximation, for the
two-particle propagators, well beyond the one-loop one used here.
Anyway, it is worth noting how this simple two-pole approximation
is able to catch the spin and charge-dominated energy regions.

\acknowledgements M.M.S. acknowledges financial support from the
\emph{Instituto Nazionale per la Fisica della Materia} and thanks
the members of the Dipartimento di Fisica ``E.R.~Caianiello'',
Universit\`{a} degli Studi di Salerno for their kind hospitality.
The authors are grateful to R. M\"{u}nzner for valuable comments
and discussions. Special thanks go to A. Kl\"umper et
al.\cite{Juttner:98} for providing us with the \emph{qtm} results.

\appendix

\section{The 2-pole Approximation}

\label{two-pole}

The doublet field (\ref{Basis}) satisfies the Heisenberg equation
\begin{equation}
i\frac{\partial}{\partial t}\Psi(i)=J(i)=\left(
\begin{array}
[c]{l}
-\mu\,\xi(i)-4t\left[  c^{\alpha}(i)+\pi(i)\right] \\
-(\mu-U)\eta(i)+4t\,\pi(i)
\end{array}
\right) \label{A1}
\end{equation}
where $\pi(i)$ is the composite field
\begin{equation}
\pi(i)=\frac{1}{2}\sigma^{\mu}\,n_{\mu}\,c^{\alpha}(i)+c(i)\left[
c^{\dagger\alpha}(i)\,c(i)\right]
\end{equation}

In the two-pole approximation we linearize the equation of motion
(\ref{A1}) as
\begin{equation}
i\frac{\partial}{\partial
t}\Psi(i)\cong\sum_{j}\varepsilon(i,j)\,\Psi(j)
\end{equation}
where the energy matrix $\varepsilon(i,j)$ is defined by
\begin{equation}
\varepsilon(i,j)=\sum_{l}\left\langle
\left\{J(i),\Psi^{\dagger}(l)\right\} \right\rangle \left\langle
\left\{\Psi(l),\Psi^{\dagger}(j)\right\} \right\rangle ^{-1}
\end{equation}

In the two-pole approximation, by assuming translational
invariance, the thermal retarded Green's function $S\left(
k,\omega\right)  ={\cal F} \left[ \left\langle {\cal R }\left[
\Psi(i)\,\Psi^{\dagger}(j)\right] \right\rangle \right] $ is given
by
\begin{equation}
S\left(  k,\omega\right)  =\frac1{\omega-\varepsilon(k)}I(k)
\end{equation}
where $\varepsilon(k)={\cal F}\left[  \varepsilon(i,j)\right]  $
is the energy matrix in momentum space and $I(k)={\cal
F}\left\langle \left\{\Psi(i),\Psi^{\dagger}(j)\right\}
\right\rangle $ is the normalization matrix.

\bibliographystyle{apsrev}
\bibliography{Biblio}

\end{document}